\documentclass[reprint, 
superscriptaddress,
amsmath,amssymb,
aps, pre,
]{revtex4-2}

\usepackage[english]{babel}
\newtheorem{theorem}{Theorem}
\usepackage{CJKutf8}
\usepackage{caption}
\usepackage{subcaption}

\usepackage{graphicx}
\usepackage{dcolumn}
\usepackage{bm}

\usepackage{soul}
\usepackage{xcolor}

\begin{document}

\begin{CJK*}{UTF8}{gbsn}
\title{Escape over a saddle by coloured noise: \\theory and numerics}%

\author{Jiayao Shao (邵家瑶)}
 \email{Contact author: jiayao.shao@warwick.ac.uk} 
\author{Tobias Grafke}
 \email{T.Grafke@warwick.ac.uk}
\author{Robert S. MacKay}
 \email{R.S.MacKay@warwick.ac.uk}
\affiliation{Mathematics Institute, University of Warwick, Coventry CV4 7AL, United Kingdom}

\begin{abstract}
Stochastic dynamical systems allow modelling of transitions induced by random disturbances, in particular from an attracting equilibrium and crossing the stable manifold of a saddle. While the small-noise limit is well-described by the large deviation principle, existing computational methods often struggle with stochastic forcings any more complicated than non-degenerate Gaussian white noise, and with unbounded time-intervals.  The primary innovations of this work are extending the framework to cater for coloured and degenerate forcing and unbounded time-horizons — scenarios that are physically realistic but numerically challenging. We cater for degenerate noise by using the Hamiltonian optimal control method.  We cater for a class of coloured noises by using linear filters on white noise. We cater for infinite time-horizon by introducing the Method of Division (MOD), a novel computational approach for approximating rare transition events, including their most likely paths and the exponential scaling laws of their transition rates. The effectiveness of MOD and the above approaches to coloured noise, is demonstrated by illustration on two examples: an inverted double-well potential and a simplified roll–heave model for ship capsize.
\end{abstract}


\maketitle
\end{CJK*}

\section{Introduction}
Dynamical systems theory is widely used in fields ranging from physics and chemistry to biology, where in particular it provides a framework for modelling state transitions by representation of motion in phase spaces. The point in phase space completely characterises the system's current state, so that the physical mechanism of a transition can be understood from its phase space trajectory. \\

A key topic within this field is the departure from a stable equilibrium under deterministic external disturbances, a phenomenon observed across a wide range of systems. Examples include the isomerisation of molecular clusters \cite{isomerisation} and the dynamic behaviour of microbeams under electrostatic actuation \cite{Krylov2010}. Classical nonlinear oscillators under deterministic forcing have long served as canonical models for studying such escape phenomena \citep{GuckenheimerHolmes1983,NayfehMook1979}. A particularly important application arises in marine engineering: ship capsize problems mentioned in \citet{Naik, suppression}, are interpreted as escapes from basins of attraction, under periodic forcing. However, real forcing for a ship is not periodic.  One can treat the effects of general bounded aperiodic forcing by extension of the theory of non-autonomous hyperbolic dynamics \cite{McSweeney}. Within this deterministic context, the transition is often conceptualized via Transition State Theory (TST) \cite{TST1938}, which identifies a critical dividing surface—typically associated with a saddle point—that trajectories must cross to escape a potential well. \\

For many purposes, however, it is better to view the forcing as coming from a stochastic process, in particular, if one wishes to determine the scaling of the transition rates with the strength of the noise. \\

From the probabilistic viewpoint, disturbances are treated as random noise and methods are employed  for stochastic differential equations. The study of noise-induced transitions traces back to \citet{Kramers}, who provided a stochastic generalisation of TST to account for the influence of thermal fluctuations and friction on escape rates. This refinement extends the purely geometric view of the transition to a dynamic one, where the `bottleneck' is crossed via noise-induced fluctuations. Subsequently, a more general and rigorous foundation was established through the large-deviation framework from \citet{bookDS}. This framework provides a systematic approach for computing the exponential scaling laws of transition rates and has found wide applicability across various disciplines, ranging from physics (e.g.,~\citet{Kautz2}) to the study of metastable systems (e.g.,~in computational neuroscience~\cite{Rabinovich2008}, atmospheric sciences~\cite{Simonnet2021}, nonequilibrium phase transitions and active matter~\cite{Grafke2017}, and climate tipping points~\cite{Soons2025}). The theory was later extended to higher-dimensional systems and non-Markovian perturbations such as coloured noise \cite{Reaction-rate}. Beyond transition rates, more recent work has focused on computing the most probable transition paths. Numerical schemes such as the minimum action method and its variants \cite{MAME,MAM} enable efficient computation of these optimal paths in many contexts, with applications in fields like hydrodynamics \cite{Hydrodynamics}. \\

Based on deterministic and probabilistic approaches, this paper introduces a novel method for analysing transitions from a stable equilibrium (for example a well in a potential landscape) to a saddle point. Our approach offers several key advantages: (i) it accommodates infinite transition time both analytically and computationally; (ii) it allows for degenerate noise (i.e.~that does not act in all directions in the phase space); and (iii) it allows the external perturbation to be coloured noise rather than white noise. Taken together, these three points result in an approach that is capable of handling physically relevant situations that are out of reach for previous methods. In safety and reliability engineering, one is generally interested in the rate of occurrence of very unlikely transitions, which can be computed by estimating the barrier height to cross in the infinite-time limit. Crucially, these infinite-time trajectories are not merely theoretical constructs. They are essential for practical estimation of mean first passage times and transition rates per unit time. For instance in the ship model, these metrics capture the dominant exponential contribution to the capsize rate, providing a rigorous basis for long-term safety assessment. Existing methods to compute barrier heights on an infinite time-horizon e.g.~\citet{longtermLDP,VE_leastaction}, rely on the invertibility of the noise covariance matrix, and thus do not apply to the degenerate noise case. These issues and the solutions provided by the Hamiltonian Optimal Control method are discussed in detail in Section \ref{HOC}. \\

The paper contributes to the goal of \citet{bujorianu2021stochasticframeworkship}, namely to integrate approaches from stochastic analysis and from non-autonomous dynamical systems theory to the question of ship capsize. It can be viewed as the stochastic companion to \citet{McSweeney}, which presents a development of the non-autonomous dynamical systems approach.\\

The outline of this paper is as follows. Section 2 explains the benefits and challenges of incorporating degenerate and filtered noise, particularly in the computation of the Freidlin-Wentzell action. Here, we apply the Hamiltonian Optimal Control (HOC) method based on \citet{MAM} to address singularities arising from noise degeneracy. A new method called the Method of Division is presented in Section 3, to allow numerical computation of infinite-time trajectories. On top of HOC, this approach isolates the two ends of the transition trajectory and linearises the dynamics around each. In particular, it separately analyses motion in the stable and unstable directions at the saddle point. This greatly enhances computational performance and convergence efficiency by concentrating computational resources on the transition region. The linearised dynamics is extended over infinite time thus the combined trajectory is traversed in infinite time. Section 4 applies the Method of Division to two case studies:~an inverted double-well model and a ship capsize model, illustrating its practical effectiveness and applicability. \\

\section{Stochastic Dynamical Systems}
A stochastic dynamical system introduces noise (usually Gaussian white) into a deterministic dynamical system, such that the system’s behaviour becomes random. This randomness can lead to phenomena like transitions from one potential well to another one, representing escapes from stable states due to stochastic effects. Many physical dynamical systems follow Newton’s Second Law, leading to a phase-space formulation where variables are partitioned into configuration and momentum coordinates. When stochasticity is introduced, external disturbances typically represent physical forces. Consequently, the noise should enter the system only through the equations for the momentum variables, leaving the configuration variables purely deterministic. This structural constraint is more realistic but leads to problems. Consider a generic stochastic differential equation (SDE) $$dX=b(X)dt+\sigma dW,$$ where $X\in \mathbb{R}^n$ and $dW \in \mathbb{R}^m$. If noise is not applied across all dimensions, the covariance matrix $C=\sigma\sigma^T$ becomes singular (non-invertible), a condition known as degenerate noise. \\

\begin{figure*}
    \centering
    \includegraphics[width=0.7\linewidth]{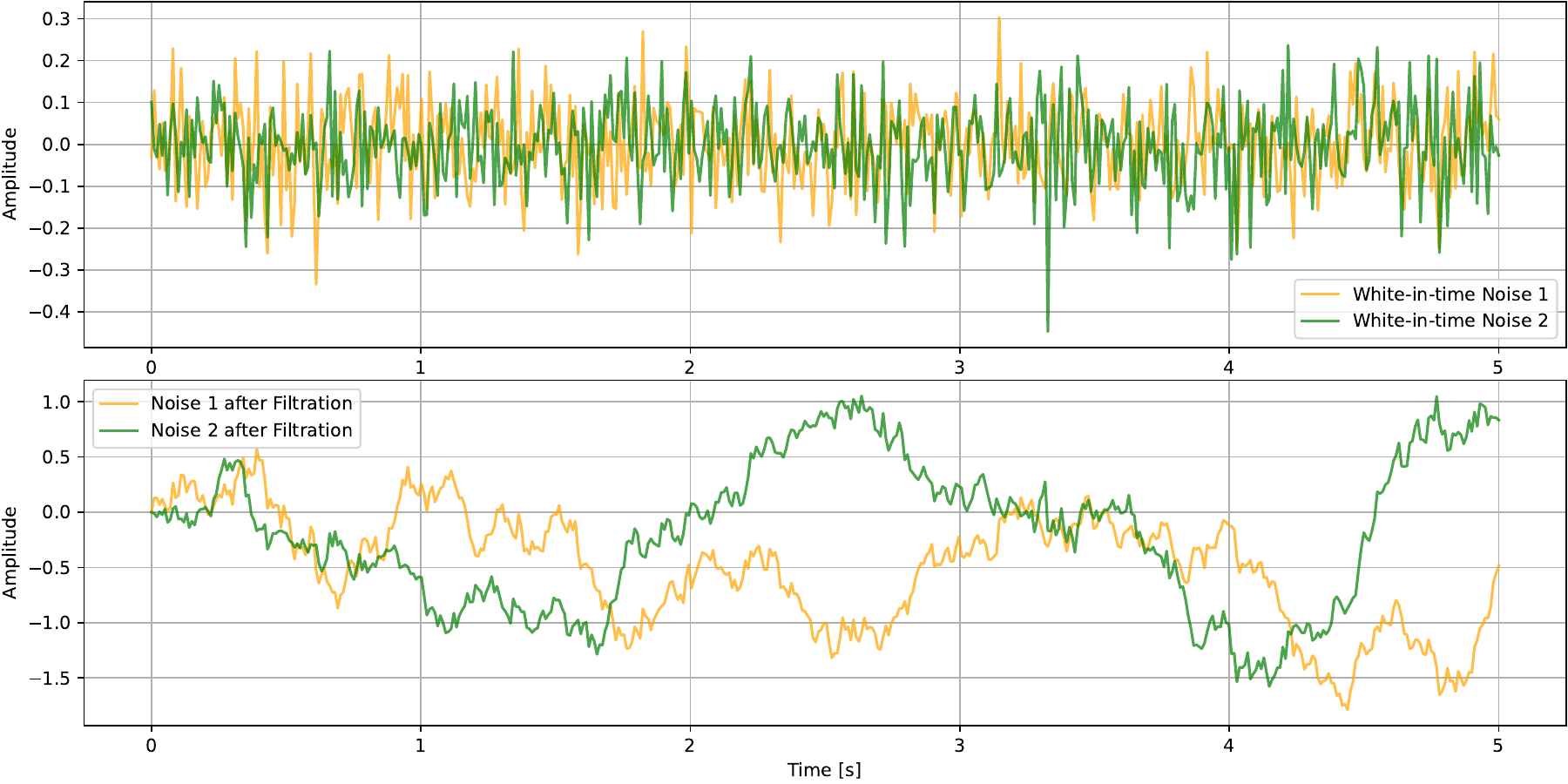}
    \caption{\raggedright Comparison between white noise and filtered noise in two-dimensional case. The top row shows increments $dW_1$ and $dW_2$ of independent white Gaussian noises. The lower row shows filtered noises $\xi_1$ and $\xi_2$ with $d\xi_1 = (-0.1\xi_1 +\xi_2)dt + dW_1$ and $d\xi_2 = (-\xi_1 -0.1\xi_2)dt + dW_2$. }
    \label{fig:noisefiltered}
\end{figure*}

Degeneracy of the noise can arise in a second way. To enhance model realism, coloured noise can be applied to the momentum dimensions, instead of white noise. A mathematically simple form of coloured noise is given by filtering white noise, i.e.~using the output $\xi$ of a stochastic differential equation of the form 
\begin{align*}
    d\xi = A\xi\, dt + \sigma \, dW,
\end{align*}
a multi-dimensional form of an equation
originally introduced by \citet{OUprocess}. In this formulation, $A$ is a general stable matrix, meaning all its eigenvalues have negative real parts. The autocorrelation properties of higher-dimensional Ornstein-Uhlenbeck (OU) processes were analysed in \citet{Stochastic_processesBook}, which is essential for modelling mechanical systems with memory. \citet{StochasticMethodsBook} further discussed the significance of filtered noise versus white noise in physical interpretations of SDEs. Adjusting the matrix $A$ allows to explicitly choose correlation times and oscillation behaviour of the random noise components $\xi$. While this attaches physical scales to the forcing, it is important to note that the filtered noise paths are continuous but nowhere differentiable with probability one \cite{adler1981geometry}. Figure \ref{fig:noisefiltered} provides the simplest example of passing white noise through a first-order low-pass filter. Note that it is impossible to plot a typical sample from white noise, because its value is nowhere defined, but what is plotted here is a sample of a discrete-time approximation. \\

Thus, a simple way to model the effect of coloured noise is to augment the state space by a set of filter variables $\xi$ with OU dynamics. It has the advantage of keeping the model in the class of standard SDEs.  But a clear consequence of such modelling is that the white noise acts only on the filter variables, not the original system, so the noise is automatically degenerate.  Note, however, that because the filter dynamics are linear, the resulting filtered noise is still Gaussian. \\

As discussed in the previous section, a common phenomenon in various domains is the occurrence of escape from a stable equilibrium, due to forcing. In this article, we will investigate a specific circumstance:~transition from a stable equilibrium to a nearby saddle, under filtered noise. The motivation is transition to the region on the other side of the saddle's stable manifold, which is assumed to be of codimension one, but for small noise, the most likely route to the other side is via a set of paths that concentrate on paths that go to the saddle. In the limit of small noise, such transitions become rare and are difficult to observe via direct sampling, but their probability can be expressed by Large Deviation Theory. \\

\subsection{Large Deviation Theory}
The theory of large deviations in probability theory deals with the behaviour of extreme tails of sequences of probability distributions, going beyond the domain of applicability of  central limit theory. It formalises the concepts of concentration of measures and extends the notion of convergence of probability measures \cite{varadhanbook}. Here we follow the scheme from \cite{MAM}. \\

Given a generic SDE taking the form 
\begin{align*}
    dX_t = b(X_t)dt + \sqrt{\epsilon}\sigma(X_{t})dW_t \quad \textrm{,} \quad X_t \in \mathbb{R}^n
\end{align*}
with a noise strength $\epsilon$,initial condition $X_0$, total time $T$ and deterministic drift $b(X_t)$, the probability that a sample path $\{X_{t}\}^T_{t=0}$ is located inside a cylinder of small radius $\delta$ around a chosen path $\phi(t)$ is said to satisfy a large deviation principle (LDP) if
\begin{align}
\label{equLDP}
    \mathbb{P}(\sup_{t \in [0,T]} |X_t -\phi(t)|<\delta) \asymp e^{-\frac{S[\phi]}{\epsilon}}
\end{align}
when $\epsilon \rightarrow 0$, where $S[\phi]$ is a functional called a `rate function'. The $\asymp$ denotes that the asymptotic behaviour of both sides is equivalent when taking the logarithm:
\begin{align*}
    \lim_{\epsilon \rightarrow 0}\epsilon\log\mathbb{P}(\sup_{t \in [0,T]} |X_t -\phi(t)|<\delta)=-S[\phi].
\end{align*}
We can write this as
$\mathbb{P}(\sup_{t \in [0,T]} |X_t -\phi(t)|<\delta) = K(\epsilon)\exp(-S[\phi]/\epsilon)$ with $K$ being some prefactor that is subexponential. The rate function was derived by \citet{bookDS} for the case of non-degenerate noise, i.e.~$C=\sigma \sigma^T$ positive-definite.
\begin{theorem}[Freidlin-Wentzell Theorem]
The sample paths of the stochastic differential equation
\begin{align*}
    dX_{t} = b(X_t)dt + \sqrt{\epsilon}\sigma(X_{t})dW_t ,
\end{align*} 
with $X_{t} \in \mathbb{R}^n$, $t \in [0,T]$ satisfy a large deviation principle with rate functional
\begin{align}
\label{action}
    S[\phi] = \frac{1}{2}\int_{0}^{T}\langle\dot \phi - b(\phi), C^{-1}(\dot \phi - b(\phi))\rangle~dt ,
\end{align}
\end{theorem}
where $C(X_{t})=\sigma(X_{t})\sigma^{T}(X_{t})$ denotes the positive-definite covariance matrix. $S(\phi)$ can be interpreted as the ``action'' for the Euler-Lagrange dynamics of (\ref{action}), and
$\phi \in \mathbb{R}^n$ is any path to which an action value can be assigned.  \\

The stated theorem provides a LDP for paths. We are interested in probability of transition, regardless of the path. The probability to hit (close to) the saddle also fulfils an LDP via  a ``contraction principle'' \cite{contracting}, where the rate function is given by the minimum over all paths that reach the saddle. The only required ingredient is that the mapping from path to endpoint is continuous (which it is), the rest follows directly via contraction. \\

As a consequence, LDT replaces the ineffective rare event sampling problem with a deterministic optimisation problem to compute the minimum action value $S^*$. Under the assumption of a good rate function \cite{bookDS}, this minimum is attained by at least one optimal transition path $\phi^*$ (though not necessarily unique). This reframing has multiple advantages: (i) the computational cost is roughly independent of the rareness of the event, whereas direct sampling of rare events becomes typically exponentially harder in the tails, (ii) the action at the minimum allows to quantify the tail scaling of the expected waiting time to observe the event, and its explicit dependence on the noise strength is known, and (iii) the minimising trajectory $\phi^*$ (possibly non-unique) represents the most likely way the rare transition is realized, allowing to identify physical mechanisms responsible for, or involved in, the transition. \\

When the noise is degenerate, however, the classical expression for the action functional (\ref{action}) becomes formally ill-posed due to the singularity of the noise covariance matrix. To extend the Large Deviation Principle (LDP) to such degenerate settings, \citet{azencott1980} introduced the use of H\"ormander's condition, ensuring that the system is hypoelliptic so that the noise propagates through the deterministic coupling. This theoretical framework was later formalized and refined by \citet{deuschel1989large}. \\

In this work, rather than relying on the probabilistic derivation of LDP, we adopt the Hamiltonian Optimal Control approach. This method provides an equivalent characterisation of the rate functional by identifying the minimum energy required to drive the system along a specific trajectory. By reformulating the problem as a deterministic variational task in more dimensions, we avoid inversion of the noise covariance matrix, thereby naturally accommodating the degeneracy of the noise. This approach does not require explicit pre-conditioning of the system and has broader applicability, extending beyond strictly hypoelliptic systems.\\

\subsection{Hamiltonian Optimal Control Method}
\label{HOC}
Minimising $S(\phi)$ is equivalent to an optimal control problem. We will first introduce the general optimal control result for a system of ODEs with non-degenerate forcing. This reproduces the same expression as the rate functional (\ref{action}). Then we discuss the Hamiltonian optimal control method, which introduces a conjugate variable for each component of the state of the system, and has no requirement for inverting the covariance matrix. This section builds upon the MSRJD formalism \cite{Martin1973, Janssen1976, DeDominicis1976}, while adopting a classical mechanics perspective as developed by \citet{Graham1985}.\\

For any trajectory $\phi$ of the dynamics $\dot\phi=b(\phi)+\sigma\eta$ defined on $t\in[0,T]$, where $b(\phi)$ is a deterministic vector field in $n$ dimensions, $\sigma$ is invertible and $\eta$ is a square-integrable forcing function, we wish to minimise its action $\frac{1}{2}\int_{0}^T|\eta|^2 dt$ subject to the boundary conditions $\phi(0)=\phi_0$ and $\phi(T)=\phi_T$. This can be written as
\begin{align*}
    \frac{1}{2}\int_{0}^T|\eta|^2~dt &= \frac{1}{2}\int_{0}^T\langle \sigma^{-1}(\dot\phi-b(\phi)),\sigma^{-1}(\dot\phi-b(\phi))\rangle~dt \\
    &= \frac{1}{2}\int_{0}^T(\dot\phi-b(\phi))^TC^{-1}(\dot\phi-b(\phi))~dt ,
\end{align*}
where $C=\sigma\sigma^T$. This expression yields the Freidlin-Wentzell theorem for its rate functional and corresponds to the standard optimal control approach for Lagrangian 
\begin{align*}
    \mathcal{L}(\phi,\dot\phi)=\frac{1}{2}\langle\dot\phi-b(\phi),C^{-1}(\dot\phi-b(\phi))\rangle.
\end{align*}
A necessary condition for an optimal solution is that it satisfy the Euler-Lagrange equations for this Lagrangian.\\

The above derivation holds only under the condition that the covariance matrix $C$ is invertible. However, in many contexts, this does not hold, as in the example we mentioned above. In such cases, the Hamiltonian Optimal Control method (which goes back to Pontryagin and is for example outlined in \cite{MAM}) provides a solution to address the issue of non-invertibility. We would like to minimise $\frac{1}{2}\int_0^T|\eta|^2dt$ subject to the constraint $\dot\phi=b(\phi)+\sigma\eta$ and the boundary conditions $\phi(0)=\phi_0$ and $\phi(T)=\phi_T$. By introducing a Lagrange multiplier function $\mu$ and vector $\beta$ (both of dimension $n$), followed by integration by parts, the objective function can be chosen to be
\begin{align*}
   J = & \frac{1}{2}\int_0^T\langle\eta,\eta\rangle~dt + \int_0^T\langle\mu,\dot\phi-b(\phi)-\sigma\eta\rangle~dt\\& +\langle\beta,(\phi(T)-\phi_T)\rangle \\
    = & \frac{1}{2}\int_0^T(\eta^T\eta-\dot\mu^T\phi-\mu^Tb(\phi)-\mu^T\sigma\eta)~dt \\
    & +\mu(T)^T\phi(T)-\mu(0)^T\phi_0 + \beta^T(\phi(T)-\phi_T) .
\end{align*}
Varying with respect to $\eta$ and $\phi$ gives the following necessary conditions for finding stationary points of the above optimisation problem: 
\begin{align*}
    \frac{\delta J}{\delta\eta} &= \eta^T-\mu^T\sigma = 0,  \\
    \frac{\delta J}{\delta\phi} &= \dot\mu^T+\mu^T\nabla b(\phi) = 0, \\
    \frac{\delta J}{\delta\phi(T)} &= \mu(T)+\beta^T = 0.
\end{align*}
Assume the objective function $J$ has a unique minimiser $\eta^*$ with the corresponding Lagrange multiplier $\mu^*$. Then, at the optimum the first equation becomes $\eta^*=\sigma^T\mu^*$. With $C=\sigma\sigma^T$ we obtain the equations for the optimised trajectory $\phi^*$ and its adjoint $\mu^*$:
\begin{subequations}
\label{instanton_eqs}
\begin{align}
    \dot\phi^* &= b(\phi^*) + C\mu^* = \nabla_{\mu}\mathcal{H}(\phi^*,\mu^*) \label{forinstan}, \\
    \dot\mu^* &= -(\nabla b(\phi^*))^T \mu^* = -\nabla_{\phi}\mathcal{H}(\phi^*,\mu^*) \label{backinstan}.
\end{align}
\end{subequations}
This pair of equations can be viewed as the Hamilton's equations of motion for
\begin{align*}
    \mathcal{H}(\phi,\mu)=\langle b(\phi),\mu\rangle+\frac{1}{2}\langle\mu,C\mu\rangle ,
\end{align*}
which is the Legendre-Fenchel transform of the Lagrangian $\mathcal{L}(\phi,\dot\phi)$ with respect to $\dot\phi$. One can solve (\ref{forinstan}) forward in time from $\phi(0)=\phi_0$ and solve (\ref{backinstan}) backward in time from $\mu(T)=-\beta$, and hope to vary $\beta$ until the desired final condition $\phi(T)=\phi_T$ is obtained. In our case the endpoint conditions could be the stable equilibrium ($\phi_0$) and the saddle ($\phi_T$) correspondingly, or points near these, as we will be interested in the limiting case of infinite time from sink to saddle. Moreover, by the forward equation (\ref{forinstan}), the action takes a simpler form: 
\begin{align}
    S^* =\frac{1}{2}\int_{0}^{T} \langle C \mu^*,\mu^*\rangle~dt.
    \label{eq:action}
\end{align}
With this reformulation, the inverse of $C$ is no longer required and the problem is well-defined even if $\sigma$ is degenerate, namely, to minimise (\ref{eq:action}) subject to (\ref{forinstan}) and end conditions. \\

The above approach can be further improved by applying the Augmented Lagrangian method \cite{Multiplier1969}. The point is that finding the right value of $\beta$ is not always easy and it is better to convexify the objective function. This method is based on the Lagrange Multiplier concept but replaces the above constrained optimisation problem with a series of unconstrained problems, by adding a penalty term to the objective function. In our notation, this gives
\begin{align*}
    \frac{1}{2}\int_0^T\langle\eta,\eta\rangle~dt + \int_0^T\langle\mu,\dot\phi-b(\phi)-\sigma\eta\rangle~dt \\
    +\langle\beta,(\phi(T)-\phi_T)\rangle + \lambda||\phi(T)-\phi_T||^2.
\end{align*}
Here $\beta$ is to mimic a Lagrange multiplier and $\lambda$ is the penalty parameter. The limit of $\lambda$ to infinity would guarantee the convergence to the ‘true’ solution but an algorithm for updating $\beta$
\begin{align*}
    \beta \leftarrow \beta + \lambda\cdot c(\phi) \quad
\end{align*}
makes it not necessary to take $\lambda$ to infinity, just large enough, thereby avoiding ill-conditioning. Here $c(\phi)$ is the associated constraint on $\phi$ which equals $c(\phi)=(\phi(T)-\phi_T)$ in our example. The improved version avoids numerical instabilities and leads to strong theoretical convergence. Detailed derivation is given in Chapter 17 of \cite{NumericalOptimisation}. This method only affects the pair of Hamilton equations through the boundary condition for $\mu$ in the backward equation (\ref{backinstan}) via $\phi(T)$ and $\lambda$. \\

\section{Method of Division}
While the previous section addresses the question of how to overcome complications of degenerate and coloured noise, we now turn our attention to the problem of an infinite time-horizon. To this effect, we introduce a new method in a general context, with one illustrative example and a more realistic but still simplified real-world application discussed in the subsequent section. This approach is specifically designed for handling transitions from a stable equilibrium to a saddle point, defined to be a hyperbolic equilibrium point with one-dimensional unstable manifold.\\

We would like to determine the probability rate for transition by computing the minimum of the rate functional for $T\rightarrow\infty$. However, numerical simulation over an infinite time-interval is infeasible. Approximating it with a finite time-interval leads to very different behaviour (if the time-interval is chosen too short) or is computationally prohibitive, especially for high-dimensional systems. Also, any transition trajectory on an infinite time-interval spends almost all its time near the fixed points, and only briefly passes through the transitional region. Any equidistant discretisation of the computational time-interval thus constitutes an extremely inefficient use of computational resources, where most effort is spent on regions where the dynamics are almost linear.\\

Furthermore, considering an infinite time-horizon introduces a numerical issue from time-translation degeneracy. Since the action functional is invariant under arbitrary shifts along an infinite time-axis, any time-shift of an optimal trajectory remains optimal. Numerically, this degeneracy causes the discretised solution to slide freely along the time domain, thereby severely undermining the robustness and reproducibility of the results and leading to significant instability. \\

The proposed method, called the Method of Division, tackles the above challenges by focusing the numerics exclusively on the nonlinear transitional regime, using analytical approximations for the semi-infinite intervals near the two equilibria via their linearisations, and requiring the transitional phase to end on a specified surface. This approach eliminates unnecessary computation near the equilibria, thus enhancing performance without sacrificing accuracy. Furthermore, the approximate dynamics around the endpoints are analysed under an infinite time scale, which allows us to study the asymptotic behaviour as $\sqrt{\epsilon} \rightarrow 0$ and $T \rightarrow \infty$ in equation (\ref{equLDP}). This makes it possible to investigate the limiting characteristics of the transition, yielding more meaningful insights into rare-event dynamics. Imposing a terminal constraint removes the time-translation degeneracy.  Specifically, we constrain the trajectory to cross a predefined ellipsoid around the saddle at the end of the transitional phase. Instead of an ellipsoid, any separating surface between the two equilibria could work, provided that the optimal transition trajectory intersects it transversally at a single point. Our choice had the advantage of helping the trajectory to find the saddle\\

\begin{figure}
    \begin{center}
    \includegraphics[width=0.4\textwidth]{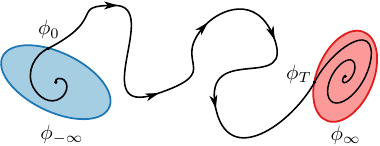}
    \caption{\raggedright A sketch of the Method of Division, where the transition process is divided into three parts: linearised dynamics around the sink $\phi_{-\infty}$ (in blue) and the saddle $\phi_{\infty}$ (in red), and the transition from $\phi_0$ to $\phi_T$.}
 \label{fig:mod-sketch}
    \end{center}
\end{figure}

The Method of Division divides the whole transition process into three parts:
\begin{enumerate}
    \item \textbf{Initial Escape ($\phi_0$)}: The process involves the system escaping from the neighbourhood of the stable equilibrium and moving to a point, $\phi_0$, near the equilibrium. This is allowed to take an infinite amount of time for interval $(-\infty,0]$.
    \item \textbf{Transitional Phase}: The core transitional phase from $\phi_0$ to somewhere on the ellipsoid around the saddle, denoted as $\phi_T$. This part occurs over the finite, effective transition time $[0,T]$.
    \item \textbf{Final Approach}: The final part in which the system moves from $\phi_T$ to the saddle. Like the initial escape, this part can also take an infinite amount of time for interval $[T,\infty)$.
\end{enumerate}
Rather than finding the optimal trajectory connecting the two equilibrium points, the method begins and terminates the optimal trajectory close enough to these points such that a linear approximation of the dynamics is accurate enough to extend it to the two equilibria. See Figure~\ref{fig:mod-sketch} as a model sketch. $\phi_{-\infty}$ and $\phi_{\infty}$ represent the sink and the saddle correspondingly. The part in blue is the initial escape and the part in red is the final approach. The trajectory in between spends a finite time $T$ for effective transition. Consequently, the total action now consists of three components: the action associated with escaping the neighbourhood of attraction, the action for the core transition as defined by the Hamiltonian Optimal Control outlined in section \ref{HOC}, and lastly the action of overcoming the unstable direction of the saddle. Importantly, the optimisation problem for the two linear parts can be solved analytically. The Method of Division handles a more difficult question: we find not only the optimal trajectory for transition but also the optimal starting point $\phi_0$ and ending point $\phi_T$. We will explain in detail the analysis for these three processes. \\

Note that by a trajectory $\phi$ we mean a solution of the dynamics
\begin{align*}
    \dot\phi = b(\phi)+\sigma\eta,
\end{align*}
where $b(\phi)$ is the deterministic vector field, and $\sigma\eta$ represents the 'forcing' that is to be controlled and optimised to give the smallest overall action. \\

\subsection{Initial Escape: Sink}
Assume without loss of generality that the sink is at the origin: $\phi_{-\infty}=\textbf{0}$. From the sink $\phi_{-\infty}$ to a nearby $\phi_0$, the dynamics can be approximated by a linearised dynamics with derivative $A_{-\infty} \in \mathbb{R}^{n \times n}$, obtained by setting $A_{-\infty}=\nabla b(\textbf{0})$. The equation of motion is then $\dot\phi = A_{-\infty}\phi + \sigma \eta$, with boundary conditions $\phi(-\infty)=0$ and $\phi(0)=\phi_0$. We can interpret the action as the energy needed to overcome friction, with external force influencing the system. Thus, the action for this part is: $\frac{1}{2}\int_{-\infty}^0 |\eta (s)|^2 ds$, subject to the constraint that this trajectory reached the point $\phi_0$. \\

The matrix $A_{-\infty}$ is assumed to be stable, i.e.~all eigenvalues have negative real parts. By Duhamel's principle (e.g.~in \citet{controltheory}),
\begin{align*}
    \phi(t)=
\int^{t}_{-\infty} e^{A_{-\infty}(t-s)}\sigma\eta(s)\, ds.
\end{align*}
In particular, we have
\begin{align*}
    \phi(0)=\int_{-\infty}^0 e^{-A_{-\infty} s}\sigma\eta(s)\, ds .
\end{align*}
To clarify the notations used: $\phi(-\infty)$ and $\phi(0)$ refer to the starting and ending points of process 1, where $\phi(-\infty)$ is assumed to be zero and $\phi(0)$ is computed using Duhamel's principle. On the other hand, $\phi_0$ represents a chosen endpoint of process 1 and is also the starting point for transitional process 2. \\

Given $\phi_0$, the endpoint  for the process 1, as a constraint, there is the following objective function by applying the method of Lagrange multipliers: 
\begin{align*}
    &F[\phi, \eta,\nu]\\&=\frac{1}{2}\int_{-\infty}^0 |\eta (s)|^2~ds + \langle\nu,(\phi_0-\phi(0))\rangle \\
    &= \frac{1}{2}\int_{-\infty}^0 |\eta (s)|^2~ds + \langle\nu,(\phi_0-\int_{-\infty}^0 e^{-A_{-\infty} s}\sigma\eta(s)~ds)\rangle \\
    &= \frac{1}{2}\int_{-\infty}^0 \eta^T \eta~ ds + \nu^T(\phi_0-\int_{-\infty}^0 e^{-A_{-\infty} s}\sigma\eta(s)~ds),\\
\end{align*}
which leads to
\begin{align*}
\frac{\delta F}{\delta \eta}&=\eta^T-\nu^T e^{-A_{-\infty} s}\sigma=0,\\
    \eta^T &=\nu^T e^{-A_{-\infty} s}\sigma.    
\end{align*}
Therefore, under the optimal trajectory and forcing:
\begin{align*}
    \phi(0) &=\int_{-\infty}^0 e^{-A_{-\infty} s}\sigma\sigma^T e^{-A_{-\infty}^T s} \nu~ds = Q\nu,\\
    \frac{1}{2}\int_{-\infty}^0 |\eta (s)|^2 ds &= \frac{1}{2}\int_{-\infty}^0 \nu^T e^{-A_{-\infty} s}\sigma\sigma^T e^{-A_{-\infty}^T s} \nu~ds \\&=\frac{1}{2}\nu^T Q \nu,
\end{align*}
where $$Q=\int_{-\infty}^0e^{-A_{-\infty} s}\sigma\sigma^T e^{-A_{-\infty}^T s}~ds.$$ 
A convenient way to compute $Q$ is to note that it satisfies the Lyapunov equation $A_{-\infty} Q+QA_{-\infty}^T=-\sigma\sigma^T$. As $A_{-\infty}$ is contracting, the solution to this Lyapunov equation exists and is unique \cite{lyapunov} and can be computed by standard routines. Furthermore, assuming controllability of the linearised dynamics around the sink ensures that $Q$ (called the controllability Gramian) is invertible \cite{controltheory}. Since at the constraint $\phi(0)=\phi_0$, we can express $\nu$ as $\nu=Q^{-1}\phi_{0}$. This leads to a further simplification of the action expression:
\begin{align*}
    \frac{1}{2}\int_{-\infty}^0 |\eta (s)|^2~ds = \frac{1}{2}\phi_0^T Q^{-1} \phi_0 .
\end{align*}

Given a chosen $\phi_0$, we have the optimal $\eta$ for the trajectory from the stable equilibrium $\phi(-\infty)$ to $\phi_0$. By substituting $s=t+\tau$, we can explicitly express $\phi(t)$ as
\begin{align}
    \phi(t) &= \int_{-\infty}^t e^{A_{-\infty}(t-s)}\sigma\eta(s)\,ds \nonumber \\
    &= e^{A_{-\infty} t}\int_{-\infty}^t e^{-A_{-\infty} s}\sigma\sigma^T e^{-A_{-\infty}^T s}\nu\, ds \nonumber \\
    &= e^{A_{-\infty} t}\int_{-\infty}^0 e^{-A_{-\infty}(t+\tau)}\sigma\sigma^T e^{-A_{-\infty}^T(t+\tau)}\nu\, d\tau  \nonumber\\
    &= \int_{-\infty}^0 e^{-A_{-\infty}\tau}\sigma\sigma^T e^{-A_{-\infty}^T\tau} e^{-A_{-\infty}^T t}\nu \,d\tau \nonumber \\
    &= Q e^{-A_{-\infty}^T t}\nu \nonumber\\
    &= Qe^{-A_{-\infty}^T t}Q^{-1}\phi_0 . \label{sinktrajectory}
\end{align}
The final equation allows for an easy sketch of the linearised dynamics. Equivalently, the optimally controlled dynamics for the linearised system around the sink is given by 
$$\dot{\phi} = -QA_{-\infty}^T Q^{-1} \phi.$$

\subsection{Final Approach: Saddle}
Similarly to the case of a sink, we can linearly approximate the dynamics around the saddle $\textbf{s}$ using a matrix $A_{\infty}$ obtained by evaluating $\nabla b(\phi)$ at the saddle $\textbf{s}$. For simplicity, let us first shift the saddle to the origin $\textbf{0}$. An equation that includes the effect of shifting the saddle will be given at the end of this subsection. The saddle has $n-1$ eigenvalues with negative real parts and one positive eigenvalue $\lambda_+>0$. \\

It is convenient to analyse the linearised system in a coordinate system that separates off the unstable direction.  Thus, we define a projection $P$ by $P=rl^T$ where $r$ and $l$ are the corresponding right and left eigenvectors of the positive (unstable) eigenvalue $\lambda_{+}$ (normalised to have $l^T r = 1$). This projection decomposes the space into $E_s \oplus E_u$, where $E_u=$ Im($P$) and $E_s =$ Ker($P$). At least one component of $l$ is non-zero, without loss of generality the first. Define a transformation matrix $U$ as
\begin{align}
\label{transformationmatrix}
    U = 
    \begin{pmatrix}
        r_1 & -\frac{l_2}{l_1} & -\frac{l_3}{l_1} & ... & -\frac{l_n}{l_1}  \\
        r_2 & 1 & 0 & ... & 0\\
        r_3 & 0 & 1 & ... & 0\\
        ... & ... & ... & ... & 0 \\
        r_n & 0 & 0& ... & 1
    \end{pmatrix} .
\end{align}
The image by $U$ of the first unit vector is $r$, and $l^TU$ is the row vector $[1,0,\ldots 0]$. This transforms $A_{\infty}$ into a matrix $\tilde{A} = U^{-1}A_{\infty} U$ taking the block form
\begin{align*}
\tilde{A} = 
\begin{pmatrix}
 \begin{array}{c|c} 
 \tilde{A_u} & 0 \\ 
 \hline 
 0 & \widetilde{A}_s 
\end{array} 
\end{pmatrix}
=
\begin{pmatrix}
  \begin{array}{c|c} 
  \lambda_+ & 0 \\ 
  \hline
  0 & \widetilde{A}_s 
\end{array} 
\end{pmatrix} ,
\end{align*}
where $\tilde{A_s} \in \mathbb{R}^{(n-1)\times(n-1)}$ is a stable matrix. Compared to diagonalisation, this transformation defined by $U$ is more efficient (it does not require computation of the other eigenvectors) and robust (diagonalisation is unstable near cases of repeated eigenvalues, and in general impossible if there are any repeated eigenvalues). There is freedom to apply any invertible coordinate change to the $(n-1)\times (n-1)$ block, if desired, but we did not use that. \\

Thus, by changing the coordinates and writing $U^{-1}(\phi-\textbf{s}) = \tilde{\phi}$, the linearised dynamics is then equivalent to
\begin{align*}
    \dot{\tilde{\phi}} = \tilde{A}\tilde{\phi} + \tilde{\sigma}\eta ,
\end{align*}
where $\tilde{\sigma} = U^{-1}\sigma$. Since $\tilde{A}$ is in block form, this can be decomposed as
\begin{align*}
    \dot{\tilde{\phi^1}} &= \lambda_{+}\tilde{\phi^1} + \tilde{\sigma}^1\eta, \\
    \dot{\tilde{\phi^i}} &= (\tilde{A_s}\tilde{\phi})^i + \tilde{\sigma}^i\eta \text{,\quad \quad \quad   for $2 \leq i \leq n$},
\end{align*}
where $\tilde{\sigma}^i$ denotes the $i^{th}$ row of $\tilde{\sigma}$. Note that, to travel from a point $\phi_T$ to the saddle, external force is only required for the $\phi^1$ component. For the components along the stable manifold, the dynamics will naturally flow toward the saddle, as long as the optimal forcing $\eta$ associated with the unstable component goes to zero.\\

Hence, we aim to find an optimal $\eta$ such that for $T\leq t\leq\infty$, the dynamics satisfies $\tilde{\phi}^1(T)=\tilde{\phi}^1_T$ and $\tilde{\phi}^1(\infty)=\textbf{0}$. Since the time interval is infinite, let us shift it to $0\leq t\leq\infty$ for simplicity in expression and calculation. The boundary condition becomes $\tilde{\phi}^1(0)=\tilde{\phi}^1_T$. By Duhamel's Principle, $\tilde{\phi^1}(0)=-\int_{0}^{\infty}e^{-\lambda_+ s}\tilde{\sigma^1}\eta(s)\, ds$. Minimising $\frac{1}{2}\int^{\infty}_0 |\eta|^2dt$ subject to the boundary conditions employing the Lagrangian Multiplier method gives objective function: 
\begin{align*}
    & G[\tilde{\phi^1_T},\eta,\beta] = \frac{1}{2}\int^{\infty}_0 |\eta(s)|^2 ds + \beta(\tilde{\phi}^1_T-\tilde{\phi^1}(0)).
\end{align*}
By similar computation as in the sink case, 
\begin{align*}
    G[\tilde{\phi^1_T},\eta,\beta] 
    &=\frac{1}{2}\int^{\infty}_0 |\eta(s)|^2ds \\&+ \beta(\tilde{\phi}^1_T+\int_{0}^{\infty}e^{-\lambda_+ s}\tilde{\sigma}^1\eta(s)ds), \\
    \frac{\delta G}{\delta\eta}&=\eta^T+\beta e^{-\lambda_+ s}\tilde{\sigma}^1 =0, \\
    \eta &= -(\tilde{\sigma}^1)^T e^{-\lambda_+ s}\beta, \\
    \beta &= \frac{\tilde{\phi}^1_T}{q}, \\
    \frac{1}{2}\int^{\infty}_0 |\eta|^2dt &= \frac{1}{2}q^{-1}(\tilde{\phi}^1_T)^2,
\end{align*}
where $q \in \mathbb{R}$ is given by
\begin{align*}
    q=\int^{\infty}_0 e^{-\lambda_{+}s}(\tilde{\sigma}^1)(\tilde{\sigma}^1)^T e^{-\lambda_{+}s}ds = \frac{(\tilde{\sigma}^1)(\tilde{\sigma}^1)^T}{2\lambda_{+}}. 
\end{align*}    
Assuming controllability of the linearised dynamics around the saddle, $q$ is non-zero. Actually, we don't need complete controllability for this, because $q$ is just the $(1,1)$-component of the controllability Gramian for infinite time. Hence, with a given $\tilde{\phi}_T$ and the optimal noise $\eta$, the controlled dynamics is
\begin{align}
\label{eq:Ahat}
  \hat{A}=\begin{pmatrix}
  \begin{array}{c|c} 
  -\lambda_+ & 0 \\ 
  \hline 
  -q^{-1}(\tilde{\sigma}^i)(\tilde{\sigma}^1)^T  & \tilde{A_s} 
\end{array} 
\end{pmatrix}
\end{align} 
where the $i^{th}$ entry of the first column is given by $-(\tilde{\sigma}^i)(\tilde{\sigma}^1)^T q^{-1}$ for $2\leq i \leq n$. $\hat{A}$ is now stable. The trajectory for the unstable component takes the form
\begin{align}
    \tilde{\phi}^1 (t) = \tilde{\phi}_T^1e^{-\lambda_{+}t},
    \label{saddleunstable}
\end{align}
and the trajectories for the stable components are 
\begin{align}
     & \tilde{\phi}^i(t) = e^{\tilde{A_s}t}\tilde{\phi}^i(0) \nonumber
     \\&+ (\tilde{A_s}+\lambda_+ I)^{-1}(e^{-\lambda_+ t}I-e^{\tilde{A_s}t})\tilde{\sigma}^i(\tilde{\sigma}^1)^T q^{-1} \tilde{\phi}^1(0)
    \label{saddlestable}
\end{align}
for $2\leq i \leq n$. These trajectories can be transformed back to the original coordinate by $\phi=U\tilde{\phi}+\textbf{s}$. Detailed computation for the analysis around the saddle is in the Appendix \ref{saddleanalysis}. \\

\subsection{Transitional phase}
After knowing a point $\phi_0$ near the sink and a point $\phi_T$ near the saddle, the Hamiltonian optimal control method allows us to find an optimal trajectory from $\phi_0$ to $\phi_T$ in time $T$, as already described. However, as discussed before, the overall problem of travelling from the sink at time $-\infty$ to the saddle at time $+\infty$ has a time-translation degeneracy and is problematic for numerical algorithms. This time symmetry issue can be removed by requiring the trajectory at a specified time to lie on a pre-defined separating surface. We choose to require $\phi(T)$ to be on an ellipsoid around the saddle, transverse to the optimum flow for reaching the saddle, defined by $\hat{A}$. One could instead require $\phi(0)$ to be on a suitable ellipsoid around the sink, or $\phi(T/2)$ to be on some surface ``half-way'' in between the sink and saddle, but we made this choice to focus the computational method on paths that approach the saddle. \\

Consider an ellipsoid centred at the saddle, defined by $(\tilde{\phi}_T)^T M (\tilde{\phi}_T) - r^2 = 0$ for some positive-definite matrix $M$, such that the optimally controlled dynamics is transverse to the ellipsoid. For $r$ small, this ellipsoid bounds a region around the saddle that is small enough for the accuracy of the linearisation approximation. The overall optimisation problem finds an optimal point on the ellipsoid to enter this domain and the linearised motion within this domain is purely analytical. To achieve the transversality, we choose the matrix $M$ to satisfy the Lyapunov equation $\hat{A}^T M + M\hat{A} + I_{n}=0$, where $I_n$ is the $n$-dimensional identity matrix, but could be any other positive-definite matrix, and $\hat{A}$ (see equation \eqref{eq:Ahat}) is the transformed linearised dynamics under control. Because $\hat{A}$ is strictly stable, the resulting $M$ is guaranteed to be positive definite. \\

It is important to distinguish between the two versions of the Lyapunov equation used for the sink and saddle analyses. Careful attention is required during implementation to ensure the correct form is applied. We provide a rigorous derivation and contextual explanation for each case in Appendix \ref{lyapunovequations}. \\

By restricting the endpoint $\phi(T)$ to this transverse ellipsoid, the continuous time-translation symmetry is successfully broken. Consequently, the non-linear optimisation phase over the finite interval $[0, T]$ becomes well-conditioned, removing the numerical ambiguity of path-sliding and allowing the optimal trajectory to be solved robustly. We have no guarantee of uniqueness, but we have made the determination of an optimal trajectory robust.\\

\subsection{Overall Problem}
We now combine the three phases into one overall problem, which involves allowing $\phi_0$ to vary near the sink, and $\Phi_T$ to vary on the chosen ellipsoid around the saddle.  The action is the sum of three parts for the three phases.  Its optimisation is carried out with respect to variations of the transition path and these endpoints.  As mentioned previously, the reason for requiring $\phi_T$ to be on the ellipsoid is to remove a time-translation degeneracy for the optimum, which would otherwise occur. \\

Firstly, the optimisation problem is to minimise the overall action functional:
\begin{align*}
   S[\phi_0,\eta] = &\frac{1}{2}\phi_0^T Q^{-1}\phi_0 + \frac{1}{2}\int_{0}^T \langle \eta,\eta \rangle\, dt \\ &+ \frac{1}{2}q^{-1}[(U^{-1}(\phi(T)-\textbf{s}))^1]^2 ,
\end{align*}
where $(U^{-1}(\phi(T)-\textbf{s}))^{1}$ represents the unstable component of the trajectory's endpoint relative to the saddle $\textbf{s}$ in the transformed coordinate system. Within this formulation, the endpoint $\phi(T)$ is uniquely determined by the choice of the initial state $\phi_0$ and the control path $\eta$. Specifically, by enforcing the initial condition constraint:
\begin{align}
\label{process1}
    &\phi_0 = \phi(0),
\end{align}
and integrating the forward equation of the Hamiltonian method:
\begin{align}
\label{process2}
    &\dot\phi = b(\phi) + \sigma\eta,
\end{align}
over the time horizon $t \in [0,T]$, the entire trajectory and thus its terminal state $\phi(T)$ is fully specified. \\

The three terms in the objective function correspond to the action accumulated across the three distinct stages defined by the Method of Division, where only the middle term is explicitly expressed as a time integral. The first term accounts for the initial escape state near the sink, entirely characterised by the optimized starting point $\phi_0$. The third term represents the remaining cost during the final approach state, where only the unstable component of the endpoint $\phi(T)$ is relevant. In the context of the Large Deviation Principle, the minimised value of $S[\phi_0,\eta]$ dictates the exponential scaling law of the transition rate. \\

In addition to the initial condition and system dynamics, a terminal constraint is imposed on the endpoint $\phi(T)$, forcing it to lie on a predefined ellipsoid surrounding the saddle point:
\begin{align}
\label{process3}
    &r^2 = \tilde{\phi}(T)^T M \tilde{\phi}(T),
\end{align}
where $\tilde{\phi}(T)=U^{-1}(\phi(T)-\textbf{s})$ is the endpoint expressed in the transformed coordinates. The optimisation problem thus reduces to finding the minimum action value by varying $\phi_0$ and $\eta$ subject to the three constraints from \eqref{process1} to \eqref{process3}.

\subsubsection{Algorithm and Numerical Solutions}
To solve this optimisation problem, we apply the Augmented Lagrangian method. The objective (cost) function is: 
\begin{align*}
    &J[\phi,\eta,\mu,\nu,\beta,\lambda] = \frac{1}{2}\phi_0^T Q^{-1}\phi_0 + \frac{1}{2}\int_{0}^T \langle \eta,\eta \rangle dt \\&+ \frac{1}{2}q^{-1}(\tilde{\phi}(T)^1)^2 
     + \langle\nu,\phi_0-\phi(0)\rangle  + \int_0^T\langle\mu,\dot\phi-b(\phi)-\sigma\eta\rangle dt \\
    & + \beta(\tilde{\phi}(T)^T M \tilde{\phi}(T) - r^2) 
    +\lambda(\tilde{\phi}(T)^T M \tilde{\phi}(T) - r^2)^2.
\end{align*}
The parameters $\nu \in \mathbb{R}^n, \mu \in \mathbb{R}^n, \beta \in \mathbb{R}$ are the Lagrange multipliers and $\lambda \in \mathbb{R}$ is a penalty parameter. Note that $\phi(0)$ and $\phi(T)$ correspond to the initial point and endpoint of the trajectory in process 2. For convenience, we can substitute $\phi(0)=\phi_1$ and $\phi(T)=\phi_N$ when discretising the trajectory $\phi$ into $N$ points. A detailed discretisation scheme is given in the Appendix \ref{discretisation}. \\

Varying the discretised cost function $J$ with respect to the variables $\phi_0$, $\eta$ and $\phi$ gives: 
\begin{align*}
    \frac{\partial J}{\partial\phi_0} &= Q^{-1}\phi_0 +\nu, \\
    \frac{\partial J}{\partial\eta_i} &= \eta_i - (\mu^T\sigma)_i, \\
    \frac{\partial J}{\partial\phi_1} &= -\nu - \mu_1 - \mu_1\nabla b(\phi_1)^T\Delta t, \\
    \frac{\partial J}{\partial\phi_i} &= \mu_{i-1} -\mu_{i} - \mu_i\nabla b(\phi_i)^T\Delta t, \\
    \frac{\partial J}{\partial\phi_N} &= \mu_{N-1} + \frac{(l^T(\phi_N-\textbf{s}))l}{q} + 2(U^{-T}MU^{-1})(\phi_N-\textbf{s})\beta \\
    &+ 4\lambda(U^{-T}MU^{-1})(\phi_N-\textbf{s})(\tilde{\phi}_N ^T M \tilde{\phi}_N -r^2)).
\end{align*}
This follows a very similar idea as discussed in the Hamiltonian Optimal Control method: the constraint and the penalty term on the ellipsoid together provide the boundary condition for a discretisation of the backward Hamiltonian equation (\ref{backinstan}) on $\mu$. The optimised noise $\eta$ can be found via gradient descent in the direction $\eta-\sigma^T\mu$ until $\eta=\sigma^T\mu$. The backward equation on $\mu$ further gives a condition on $\nu$ so that $\phi_0$ can be optimised. Hence, we have two variables to optimise: $\phi_0$ and $\eta$ where $\eta$ is a function of time $t\in[0,T]$. The trajectory $\phi$ can be computed from $\phi_0$, knowing $\eta$, through the forward Hamilton's equation (\ref{forinstan}), and the endpoint $\phi_N$ is forced on the ellipsoid that gives the smallest action. The algorithm is as follows: 
\begin{enumerate}
    \item Initialise $\phi_0$, $\eta$, the Lagrange multiplier $\beta$ and the penalty parameter $\lambda$. Choose convergence thresholds $\tau_{\text{grad}}$ and $\tau_{\beta}$.
    \item For fixed $\beta$ and $\lambda$, update $\phi_0$ and $\eta$:
    \begin{enumerate}
        \item Solve the forward Hamiltonian dynamics
        \begin{align*}
            \dot\phi = b(\phi) + \sigma\eta, \quad \phi(0) = \phi_0,
        \end{align*}
        on $[0,T]$ to obtain $\phi(T)$.
        \item Solve the backward adjoint dynamics
        \begin{align*}
            \dot\mu = - \nabla b(\phi)^{\top}\mu,
        \end{align*}
        with terminal condition
        \begin{align*}
              & \mu_{N-1} = -\frac{(l^T(\phi_N-\textbf{s}))l}{q} \\
              &- 2(U^{-T}MU^{-1})(\phi_N-\textbf{s})(\beta + 2\lambda(\tilde{\phi}_N ^T M \tilde{\phi}_N -r^2)),
        \end{align*}
        to obtain $\mu$ and $\nu$ with 
        \begin{align*}
            \nu = - \mu_1 - \mu_1\nabla b(\phi_1)^T\Delta t . 
        \end{align*}
        \item Evaluate the gradients of the augmented cost functional with respect to the optimisation arguments:
        \begin{align*}
            \nabla_{\eta}J &= \eta-\sigma^T\mu \\
            \nabla_{\phi_0}J &= Q\phi_0+\nu,
        \end{align*}
        and update $\eta$ and $\phi_0$ using a chosen optimisation algorithm.
        \item Repeat steps (a)-(c) until the norm of the overall gradient is strictly smaller than $\tau_{\text{grad}}$. 
    \end{enumerate}
    \item Update the penalty parameter and the Lagrange multiplier with some chosen constant $1<c\ll 2$:
    \begin{align*}
        \beta_{k+1} = \beta_k + \lambda_k(\tilde{\phi}(T)^\top M \tilde{\phi}(T)-r^2),
        \quad
        \lambda_{k+1} = c\lambda_k.
    \end{align*}
\end{enumerate}
\noindent Repeat step 2 and 3 until $|\beta_{k+1}-\beta_k| < \tau_\beta$ such that the constraint is satisfied, i.e. $\lambda_k(\tilde{\phi}(T)^\top M \tilde{\phi}(T)-r^2) \approx 0$. Output: Optimised $\eta$ and $\phi_0$. \\

Note that for Step 2(c) we utilised the non-linear conjugate gradient method, as it exhibits significantly faster convergence for this path-optimisation problem. A detailed discussion on the inner product definition and the resulting gradient norm is provided in Appendix \ref{gradientnorm}.

\subsection{Linearisation Robustness}
In this section, we analyse the linearisation error from the initial escape phase and the final approach phase.  It can be skipped on a first reading.  We turn on the nonlinear part of the vector field:
\begin{align*}
    b(\phi)=A\phi+\delta b(\phi),
\end{align*}
where $A$ is the linearised dynamics that could be either $A_{-\infty}$ or $A_{\infty}$. We write the perturbed trajectory as $\phi=\bar{\phi}+\delta\phi$ and the perturbed forcing as $\eta=\bar{\eta}+\delta\eta$, where $\bar{\cdot}$ represents the zero-th order term and $\delta\cdot$ is the first order correction. The optimisation problem becomes to minimise the action
\begin{align*}
    S = \frac{1}{2}\int (\bar{\eta}+\delta\eta)^T(\bar{\eta}+\delta\eta)~dt ,
\end{align*}
subject to the constraints
\begin{align*}
    \dot\phi = \dot{\bar{\phi}}+\dot{\delta\phi}=A(\bar{\phi}+\delta\phi)+\delta b(\bar{\phi}+\delta\phi)+\sigma(\bar{\eta}+\delta\eta) 
\end{align*}
and boundary conditions at the ends of the trajectory. Following a similar calculation via the Hamiltonian optimal control framework introduced earlier, where $\mu$ acts as the Lagrange multiplier, the first-order equations for the perturbations (taking the sink as an example; see Appendix \ref{linearisationerror} for details) are given by
\begin{align*}
    \dot{\delta\phi} &= A\delta\phi+\delta b(\bar{\phi})+C\delta\mu, \\
    \dot{\delta\mu} &= -A^T\delta\mu -\nabla \delta b^T\bar{\mu} ,
\end{align*}
with boundary condition $\delta\phi(-\infty)=\textbf{0}$ and $\delta\phi(0)=0$ such that $\bar{\phi}(0)+\delta\phi(0)=\phi_0$ is the boundary constraint. Similarly to the zero-th order case, we can forward integrate the $\delta\phi$ equation and use $\delta\phi(0)=0$ as the boundary condition to backward integrate the $\delta\mu$ equation. \\

After solving the $\delta\mu$ equation, we can express the first order change in the action as
\begin{align*}
    \delta S &= \int_{-\infty}^0 \bar{\eta}^T\delta\eta~dt, \\
    &= \int_{-\infty}^0 \bar{\mu}(s)^TC\delta\mu(s)~ds, \\
    &= -\bar{\phi}_0^TQ^{-1}\int_{-\infty}^0 e^{-As}\delta b(\bar{\phi}(s)) ~ds, \\
    &=- \bar{\phi}_0^TQ^{-1}\int_{-\infty}^0 e^{-As}\delta b(Qe^{-A^Ts}Q^{-1}\bar{\phi}_0) ~ds .
\end{align*}
This is of order $O(\bar{\phi}_0^3)$. By comparing it with the zero-th order action under linearisation $\bar{S} = \frac{1}{2}\bar{\phi}_0^TQ^{-1}\bar{\phi}_0$: the relative change is
\begin{align*}
    \frac{\delta S}{\bar{S}} \approx \frac{2|\delta b(\bar{\phi}_0)|}{c|\bar{\phi}_0|}
\end{align*}
where $c$ is the slowest contraction rate for $A$. We have to check that $|\delta b(\bar{\phi}_0)| \ll c|\bar{\phi}_0|$ for a good linearisation approximation. \\ 

In addition to determining an order of magnitude for the accuracy of the linear approximations, the first order correction to the action derived here from linearisation error can be added to the computed action to produce a more accurate value.

\section{Examples}
\subsection{Inverted Double Well}
Consider an inverted double well model with potential $V(x)=\frac{1}{2}x^2-\frac{1}{4}x^4$. The Newton's equation of motion for a particle with mass $m$ in this potential is $m\ddot x=x^3-x$. 
For simplicity, assume $m=1$. To illustrate the method of division, together with our way of treating filtered noise, let us make this a damped system and add the simplest filtered noise to the model:
\begin{align*}
dx &= y\,dt,\\
dy &= (x^3-x-ky+z)dt,\\
dz &= (-z)dt + \sqrt{\epsilon}\,dW,
\end{align*}
where $k$ is a positive constant denoting the damping coefficient, and $z$ is the one-dimensional filtered noise added. White noise is only added to the last line, which gives degenerate noise for the system as a whole, i.e., the covariance matrix is singular.\\

The model is unrealistic in that $V(x) \to -\infty$ as $x \to \pm \infty$ and the dynamics suffers finite-time blowup, but we are interested here just in reaching one of the two saddles ($x=\pm 1$) from the sink at $x=0$. It is easy to make more realistic models, e.g.~$V(x) = a\cos x - b\cos 2x$ with $4b>a>0$, but simpler to illustrate with a polynomial rather than a trigonometric function. \\

The above simultaneous equations could also be expressed in the generic SDE format as
\begin{align*}
    &d\phi = b(\phi)dt+\sigma dW,
\end{align*}
with
\begin{align*}
    \phi= \begin{pmatrix}
        x \\ y \\ z
    \end{pmatrix}
    , b(\phi)=\begin{pmatrix}
        y \\ x^3-x-ky+z \\ -z
    \end{pmatrix}
    \ \text{and} \ \sigma=\begin{pmatrix}
        0 \\ 0 \\ \sqrt{\epsilon}
    \end{pmatrix} .
\end{align*}
The deterministic dynamics of this system have three equilibrium points: the stable equilibrium $(0,0,0)$ and the two saddles $\pm(1,0,0)$. Due to symmetry, consider only one saddle at $(1,0,0)$. The linearised dynamics $\nabla b$ around the two critical points are given by
\begin{align}
    A_{-\infty} =
\begin{pmatrix}
    0 & 1 & 0 \\
    -1 & -k & 1\\
    0 & 0 & 1 
\end{pmatrix} ,
\quad
    A_{\infty} = 
\begin{pmatrix}
    0 & 1 & 0 \\
    2 & -k & 1\\
    0 & 0 & 1
\end{pmatrix} .
\label{equ_linearisation}
\end{align}
Choose $k<1$ such that the sink has one negative real eigenvalue and two complex eigenvalues with small negative real parts. The saddle has one positive real eigenvalue and two negative real eigenvalues. \\

\begin{figure}
    \centering
    \includegraphics[width=1.0\linewidth]{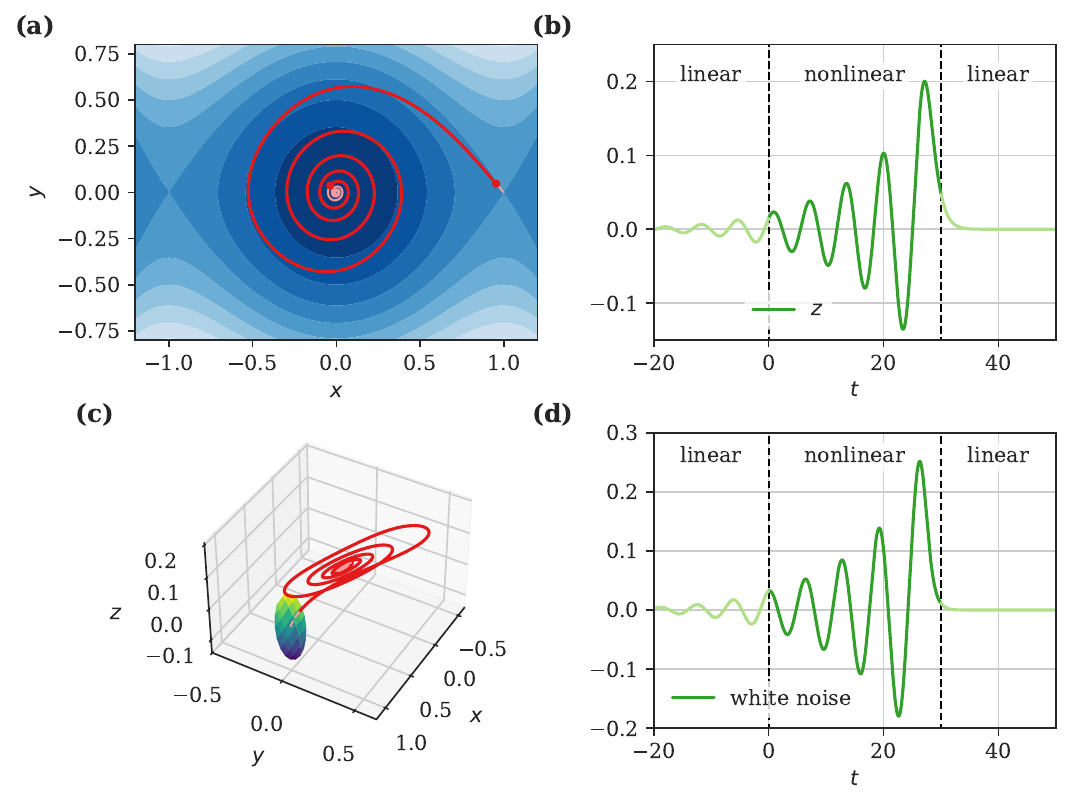}
    \caption{\raggedright \small{The optimal trajectory from the sink to the saddle in three sections: linearised dynamics at both ends and non-linear dynamics for the transition. The figures on the right show the optimal filtered noise and the corresponding realisation of the white noise.}}
    \label{fig:3Dshipmodel}
\end{figure}

The transition problem is to find the probability rate to escape over one of the saddles if the system starts at the origin, as well as the trajectory it would take. Figure \ref{fig:3Dshipmodel} illustrates the result by the Method of Division. In figure \ref{fig:3Dshipmodel}(a), the optimal trajectory is shown on a background of level sets of the Hamiltonian $H(x,y)=V(x)+\frac{y^2}{2}$ for the undamped system. Note the Hamiltonian $H(x,y)$ is in the sense of mechanics and is different from the Hamiltonian of the optimal control method mentioned before. Figure \ref{fig:3Dshipmodel}(c) is a 3D plot of the trajectory, showing also the ellipsoid defined around the saddle. The dashed pink line is the linearised trajectory within the ellipsoid. Figure~\ref{fig:3Dshipmodel}(b) and Figure~\ref{fig:3Dshipmodel}(d) show the filtered noise dimension and the corresponding Gaussian white noise. The linear part (light green) and non-linear part (dark green) are seen to join smoothly, supporting the validity of use of the linearised approximations at the two ends. The computed action value is $S =0.1330$. 

According to the large deviation principle (\ref{equLDP}), the probability rate scales asymptotically as $e^{-S/{\epsilon}}$. For instance, taking a weak disturbance with $\sqrt{\epsilon}=0.1$, the probability rate is approximately $K\times1.67\times 10^{-6}$, where $K$ denotes the prefactor in the LDP. We do not know the prefactor, but if we assume that it varies slowly with $\epsilon$ then we can deduce for example that if the noise amplitude $\sqrt{\epsilon}$ is increased from $0.1$ to $0.2$ then the transition rate increases by a factor of approximately $e^{0.133*75} \approx 21483$. In non-gradient systems, the pre-exponential factor is not merely a constant but depends on the entire history of the dynamics along the most probable trajectory \cite{maier1993escape}. But we treat it as a constant in the example section due to the significant complexity involved in its analytical determination.\\

\subsection{Ship Capsize Model}
\begin{figure}
    \centering
    \includegraphics[width=0.8\linewidth]{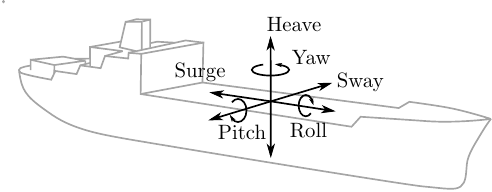}
    \caption{\small{6 degrees of freedom of a ship}}
    \label{fig:shipmodel}
\end{figure}

A more complicated example is a simple two degree of freedom ship capsize model, proposed by \citet{suppression}. This archetypal capsize equation of the ship is designed for the beam sea, where waves travel perpendicular to the body of the ship. It has two degrees of freedom:~the heave $z$ and the roll $\theta$ (see Figure \ref{fig:shipmodel}). The ship potential is given by 
\begin{align*}
    V(\theta,z) = \frac{1}{2}c\theta_{c}^2[(\frac{\theta}{\theta_{c}})^2 - \frac{1}{2}(\frac{\theta}{\theta_c})^4] + \frac{1}{2}h[z-\frac{1}{2}\gamma\theta^2]^2.
\end{align*}
The roll and heave natural frequencies of the ship are proportional to $\sqrt{c}$ and $\sqrt{h}$, with coefficients $I^{-1/2}$, $m^{-1/2}$ respectively, where $I$ is the moment of inertia for roll and $m$ is the mass of the ship. Here, $\gamma$ is a ship parameter that couples roll and heave, and $\theta_c$ denotes the static capsize angle. The restoring force in heave is $-\partial V/\partial z$ and the restoring moment in roll is $-\partial V/\partial \theta$. All variables are non-dimensionalised. Then by distinguishing the momentum dimensions and the configuration dimensions and by adding damping to heave and roll, and adding filtered noise we obtain a six-dimensional stochastic dynamical system:
\begin{align*}
    & \dot z = v_z, \\
    & \dot v_z = \frac{h}{m}(\frac{1}{2}\gamma \theta^2-z) - k_{1}v_z + B_{11}\xi_1 + B_{12}\xi_2,\\
    & \dot \theta = v_\theta, \\
    & \dot v_\theta = \frac{c}{I} \theta(\frac{\theta^2}{\theta_{c}^2}-1)+\frac{h}{I}\gamma\theta 
    (z-\frac{1}{2}\gamma \theta^2) -k_{2}v_\theta  + B_{21}\xi_1 \\&+ B_{22}\xi_2,\\
    & \dot\xi_1 = A_{11}\xi_1 + A_{12}\xi_2 + \sqrt{\epsilon}\sigma_{11}\eta_1 + \sqrt{\epsilon}\sigma_{12}\eta_2, \\
    & \dot\xi_2 = A_{21}\xi_1 + A_{22}\xi_2 + \sqrt{\epsilon}\sigma_{21}\eta_1 + \sqrt{\epsilon}\sigma_{22}\eta_2, 
\end{align*}
where $v_z$ and $v_\theta$ are the velocity coordinates. The variables $\xi_i$ are the added filtered noise and $\eta_i$ are the `derivative' of independent Wiener processes. This structure results in a system with filtered and degenerate noise, which provides a more realistic representation of ocean wave behaviour. As demonstrated by \citet{dimentberg1988statistical}, such a filtered noise approach captures parametric excitation far more accurately than a simple white noise model, as it accounts for the narrow-band spectral characteristics of the sea state. \\

Here we adopt a representative choice of the matrix $A$ that determines the spectral properties of the filtered noise:
\begin{align*}
    A=
    \begin{pmatrix}
    -\alpha & \omega \\
    -\omega & -\alpha
    \end{pmatrix}
\end{align*}
This specific form yields complex eigenvalues $-\alpha\pm i\omega$, where $\omega$ is the characteristic oscillation frequency of the noise and $\alpha$ describes the decay of the noise's autocorrelation.\\

Similarly to the inverted double-well example, the unforced system has three equilibrium points, a stable equilibrium at the origin $\textbf{0} \in \mathbb{R}^6$ and two saddle points at $(z,v_z,\theta,v_\theta,\xi_1,\xi_2)=(\frac{1}{2}\gamma \theta_c^2,0,\pm \theta_c,0,0,0)$ for some constant $\theta_c$. The saddles have a 1-dimensional unstable manifold and a 5-dimensional stable manifold in the relevant parameter range. The structure and symmetry of the saddles fit with physical intuition: the ship loses balance and capsizes when it swings too severely to the left or the right. This instability comes only from the roll direction $\theta$ and $\pm \theta_c$ is the corresponding capsize angle. \\

Note that there exists a redundancy in the coefficient matrices $B$ and $\sigma$, which can be removed by exploiting the rotation and scaling symmetries in the underlying noise spaces. Let $R \in SO(2)$ be a rotation matrix. The two-dimensional white noise $\eta$ is invariant under this orthogonal rotation. Moreover, our choice of the drift matrix $A$ commutes with such rotations (i.e., $R^{-1}AR = A$), allowing us to rotate the coloured noise coordinates $\xi$ and the driving noise $\eta$ independently. By utilising these two rotational degrees of freedom, the coupling matrix $\sigma$ can be reduced to a diagonal form with non-negative entries. The scaling freedom of the smallness parameter $\epsilon$ further allows us to normalise these entries. Therefore, the matrix $\sigma$ can be simplified to
\begin{align*}
    \sigma=
    \begin{pmatrix}
    \cos{\alpha} & 0 \\
    0 & \sin{\alpha}
    \end{pmatrix}
\end{align*}
with a single parameter $\alpha \in [0, \pi/2]$. Finally, the remaining scaling freedom in the transformation of $\xi$ allows us to normalise the matrix $B$ such that $\|B\| = 1$ under an appropriate matrix norm. These transformations effectively reduce the system to a general matrix $B$ constrained by $\|B\|=1$ and a single tuning angle $\alpha$, thereby significantly decreasing the number of free parameters.\\

\begin{figure*}
  \centering
  \begin{subfigure}{\columnwidth}
    \centering
    \includegraphics[width=0.95\linewidth]{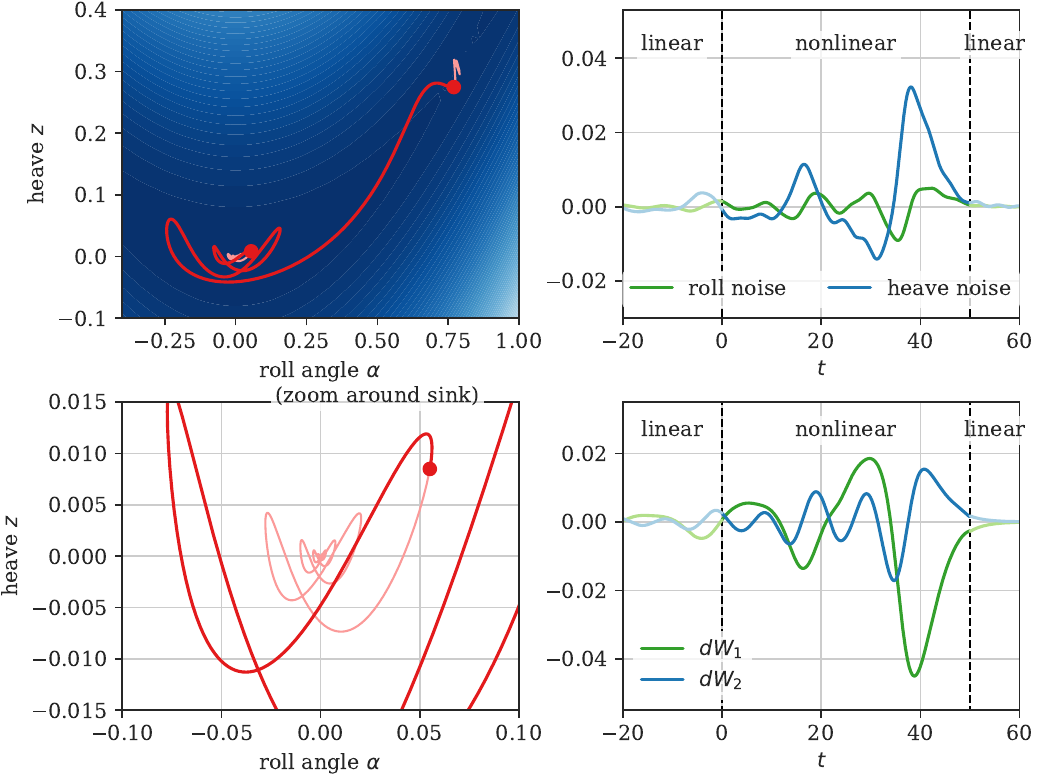}
    \caption{transition time $T=50$}
    \label{fig:T50}
  \end{subfigure}
  \begin{subfigure}{\columnwidth}
    \centering
    \includegraphics[width=0.95\linewidth]{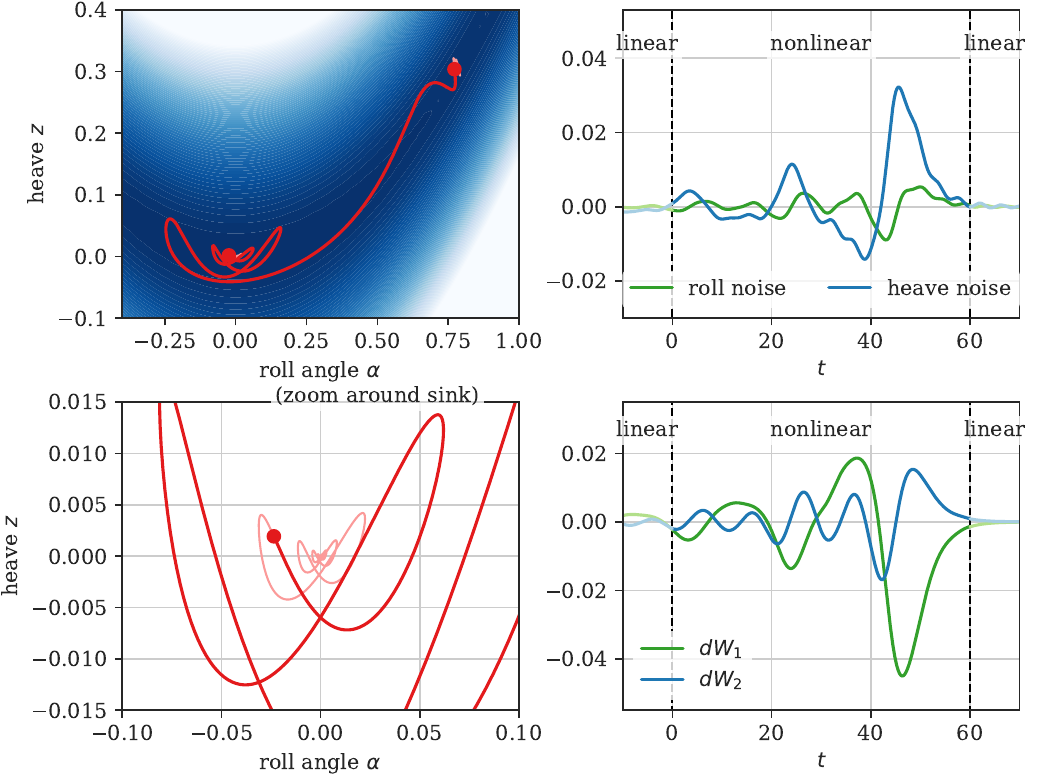} 
    \caption{transition time $T=60$}
    \label{fig:T60}
  \end{subfigure}
  \caption{\raggedright A comparison between two ship capsize trajectories and their corresponding optimal forces under different effective transition times. The forces on the filter variables are denoted as $\tilde{\eta}_i = \sigma_{i1}\eta_1 + \sigma_{i2}\eta_2$. Note that the two trajectories are virtually identical; changing $T$ merely shifts the crossover point where the dynamics transitions from the linearised to the nonlinear regime.} 
  \label{fig:timeinvariance}
\end{figure*}

Figures \ref{fig:timeinvariance} show representative results for the ship capsize problem. The specific physical parameters of the ship and noise parameters are systematically detailed in Table \ref{tab:ship_parameters}, adopting the simplified choice after eliminating the redundancies. For the chosen parameter values, we have evaluated the controllability Gramian at the sink and its unstable-unstable component at the saddle as part of our numerical implementation, confirming their invertibility and the validity of the local controllability assumptions. The red curves indicate the optimal capsize trajectory—i.e., the most likely path to capsize—while the pink curves illustrate the linearised dynamics near the equilibrium points. In both cases, the linearised dynamics align well with the nonlinear dynamics at the boundaries. The blue and green curves represent noise: filtered noise applied to the heave and roll, and white noise applied to the filter coordinates.\\

\begin{table}[htbp]
\centering
\caption{Physical and stochastic parameters utilized in the ship capsize numerical simulation.}
\label{tab:ship_parameters}
\resizebox{\columnwidth}{!}{%
\begin{tabular}{lll}
\hline
\textbf{Parameter} & \textbf{Description} & \textbf{Value} \\ \hline
$m, I$ & Ship mass and moment of inertia & $1.0, 1.0$ \\
$k_1, k_2$ & Damping coefficients (heave and roll) & $0.1, 0.1$ \\
$h, c, \gamma, \theta_c$ & ship coefficients & [0.4,0.1,1,$\pi/4$] \\
$B$ & coupling matrix & 
$\begin{pmatrix} 1/\sqrt{5} & 0 \\ 0 & 2/\sqrt{5}
\end{pmatrix}$
\\
$\sigma$ & noise matrix & $\begin{pmatrix} 1/\sqrt{2} & 0 \\ 0 & 1/\sqrt{2}
\end{pmatrix} $  \\
\hline
\end{tabular}
}%
\end{table}

In Figure \ref{fig:T50}, the effective transition time is $T=50$, whereas in Figure \ref{fig:T60} it is $T=60$, with all other parameters held fixed. The action values are identical to three significant figures in both cases: $S=0.0176$. The resulting trajectories are very similar as well, as indeed they should be; the main change as $T$ increases is that the transition from the linearised to the nonlinear regime occurs closer to the sink. This highlights the numerical advantage of the Method of Division: one can identify a sufficient transition time $T^*$ such that for any $T>T^*$, the trajectory $\phi[T]$ closely resembles $\phi[T^*]$, differing only by extending further within the linearised region near the sink. The action value also stabilises for $T>T^*$, giving the `true' solution to the problem. This saves computation cost to a great extent when investigating the change to the trajectory for longer $T$. The leading corrections from the linearisation errors were found to be less than 0.1\%, so it made little difference whether we added them in or not. \\

\subsubsection{Parameter Study}
Different parameter choices lead to distinct capsize behaviours, actions, and consequently, varying capsize rates. In this section, we examine two primary factors influencing these outcomes: the spectral properties of the external disturbances and the characteristics of the ship itself. The external disturbances—mainly waves—are modelled using filtered noise with the characteristic oscillation frequency $\omega$ (see for example Fig. \ref{fig:noisefiltered}). \\

The transition rate is typically expressed via the LDP form $e^{-S/\epsilon}$ where the action $S$ is evaluated relative to a constant white noise intensity $\epsilon$. However, when the characteristic filter frequency varies, the effective intensity of the resulting coloured noise also changes. Therefore, $\epsilon$ is no longer a suitable metric to characterise the size of the stochastic forcing. Instead of analysing the frequency-dependence of capsize rates at fixed $\epsilon$, it is more meaningful to fix a quantification of the intensity of the forcing. For this purpose, we scale the action by a factor $\lambda_{max}$ (see Appendix \ref{noiseintensity} for the definition and a detailed derivation). By comparing the normalised actions $S\cdot\lambda_{max}$ across different wave frequencies $\omega$, we effectively isolate the system's inherent dynamical sensitivity from frequency-dependent noise energy levels. This approach allows us to identify the resonance conditions under which the ship is most susceptible to capsize. Throughout this analysis, the heave and roll frequencies are kept fixed to isolate the spectral effects of the noise. On the other hand, to investigate the effect from different ship structures, we fix the filtered noise frequency and roll frequency while varying the ratio between heave and roll frequencies. \\

\begin{figure*}
    \centering
    \begin{subfigure}{0.48\textwidth}
        \centering
        \includegraphics[width=\linewidth]{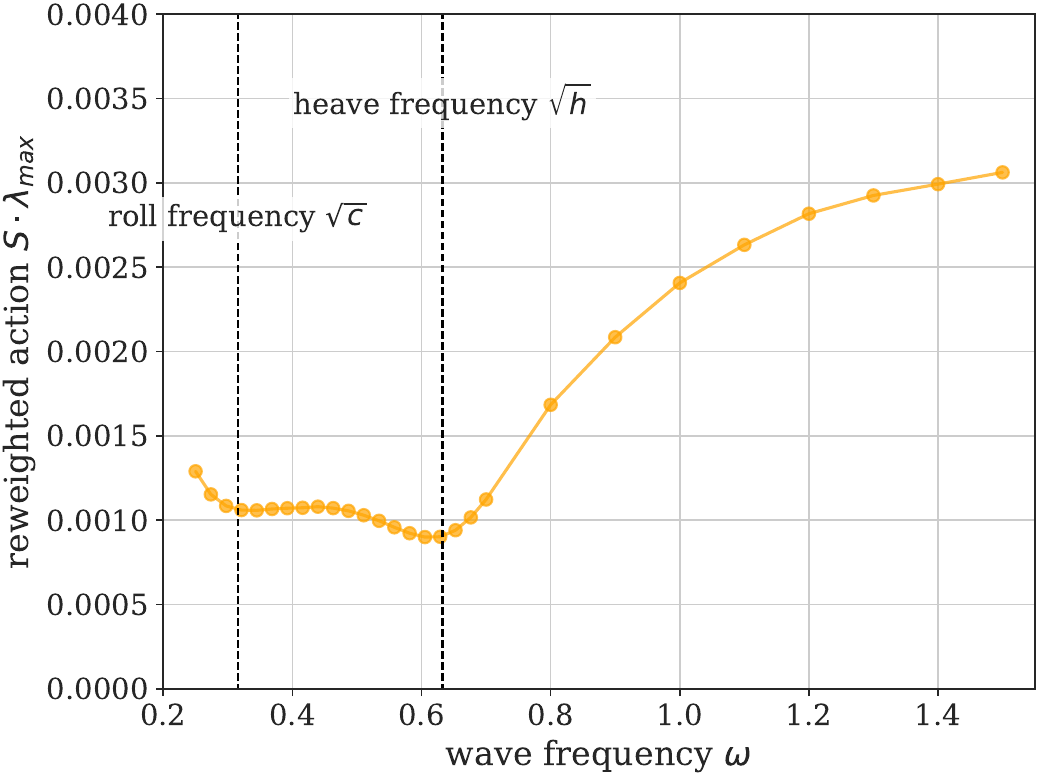}
        \caption{Varying wave frequency}
        \label{fig:wave_parameter}
    \end{subfigure}
    \begin{subfigure}{0.48\textwidth}
        \centering
        \includegraphics[width=\linewidth]{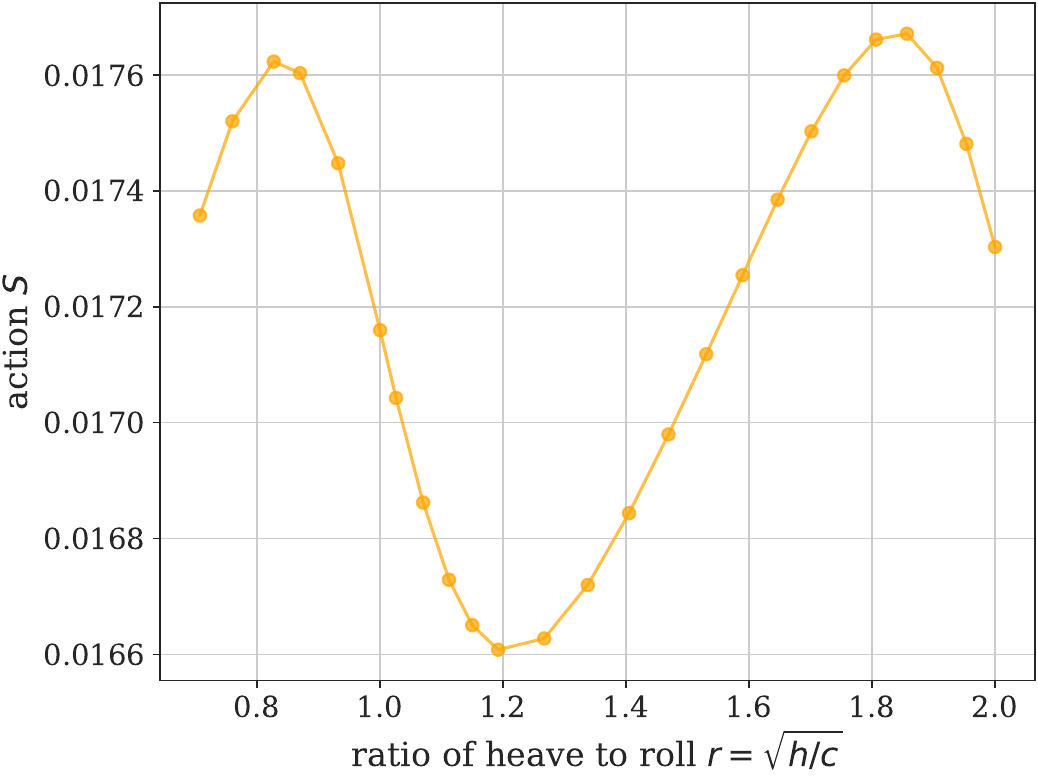}
        \caption{Varying ratio $r$ between heave and roll}
        \label{fig:ship_parameter}
    \end{subfigure}

    \captionsetup{justification=raggedright,singlelinecheck=false}
    \caption{\raggedright Parameter studies on effects of wave frequency and ship's natural frequencies on the minimal action for escape. In the case of changing wave frequency, the minimal action has been scaled to an appropriate quantification of the noise intensity.}
    \label{fig:parameter_study}
\end{figure*}

Figure \ref{fig:wave_parameter} illustrates the variation of the reweighted action $S\cdot\lambda_{max}$ over different wave frequencies, where the resonant heave and roll frequencies are marked by black dashed lines. Since a higher action value corresponds to a lower capsize probability, the local minimum observed near the heave frequency identifies the most critical frequency range for transition. In contrast, the plateau to the right of the roll frequency suggests a frequency-insensitive regime where nonlinear damping plays a dominant role. As the ship approaches roll resonance, the rapid increase in nonlinear dissipation can counter the additional energy input from waves, creating a temporary dynamic equilibrium that stabilises the action. \\

The local minimum around the heave frequency indicates a specific resonance region where the ship’s inherent dynamics synchronise with the incoming wave energy, leading to rapid energy accumulation and eventually capsize. While capsize is formally defined by the roll angle exceeding a critical threshold, our results demonstrate that resonance with heave motion is often more dangerous. This is consistent with the catastrophe model proposed by \citet{zeeman1977catastrophe}, where vertical oscillations (heaving) redistribute buoyancy and induce parametric instability in the roll equilibrium. \\

In this analysis, the heave-roll frequency ratio is kept constant at $2$, which is a typical ratio for vessels such as destroyers \citep{zeeman1977catastrophe}. This configuration corresponds to a Fermi resonance regime. So it is interesting that the wave frequency for minimum weighted action coincides precisely with the heave frequency, confirming that heave oscillations can dominantly destabilise the ship and initiate severe rolling. Thus, the synchronisation of external wave forcing with the heave mode represents the most susceptible state for the system.\\

Conversely, at higher wave frequencies (the right end of the plot), the ship’s relatively longer natural period makes it less responsive to the rapid wave excitations. In this regime, the energy transfer from the waves to the roll motion becomes highly inefficient, resulting in high stability. \\

Figure \ref{fig:ship_parameter} illustrates the action values under different ratios between heave and roll frequencies, defined as $r=\sqrt{h/c}$ (recall that the roll and heave frequencies are proportional to $\sqrt{c}$ and $\sqrt{h}$, respectively). In this experiment, the characteristic frequency of filtered noise is held constant at $1.5$ and the roll frequency is fixed at $1$. By varying the ratio $r$ from $0.7$ to $2$, we effectively span the most dynamically sensitive range, including both the simple resonance ($r=1$) and the Fermi resonance ($r=2$). \\

The action attains a minimum of approximately $0.0166$ at $r \approx 1.2$, whereas it reaches its maximum value of roughly $0.0177$ at $r \approx 1.85$. Since the probability rate follows the asymptotic scaling $\mathbb{P}\sim Ke^{-S/\epsilon}$ where the prefactor $K$ depends subexponentially on the system parameters and noise parameter $\epsilon$, we can estimate the relative likelihoods by assuming $K$ remains constant between these two cases. The ratio of probability rates for capsize is then expressed as
\begin{align*}
    \frac{\mathbb{P}(\text{capsize rate when $r\approx1.2$})}{\mathbb{P}(\text{capsize rate when $r\approx1.85$})} = e^{-\Delta S/\epsilon} 
\end{align*}
where $\Delta S \approx 0.0011$ represents the action discrepancy. Due to the exponential dependence on $\epsilon$, small changes in the noise intensity can lead to orders-of-magnitude changes in the relative likelihood of capsizing. For instance, at $\epsilon = 10^{-2}$, the ship is approximately $1.12$ times more likely to capsize at $r \approx 1.2$ than at $r = 1.85$; at $\epsilon = 10^{-3}$, this factor increases to roughly $3$; and at $\epsilon = 10^{-4}$, the probability increases drastically, reaching a likelihood approximately $6 \times 10^4$ times higher. These findings are consistent with results of \cite{suppression}, which suggested that heave-to-roll frequency ratio near 2 reduced susceptibility to capsize, in contrast to expectations based on parametric resonance. \\

\section{Conclusion}
In this paper, we introduced the Method of Division to address transition problems in stochastic dynamical systems over an unbounded time-scale and demonstrated its effectiveness through two examples:~an inverted double well and a ship capsize problem. The proposed method accommodates both degenerate and filtered noise, making it more adaptable to real-world scenarios. It also enables an infinite time scale by focusing computational effort on the nonlinear transitional phase. We illustrated the method with a transition from a stable sink to a saddle with a one-dimensional unstable manifold and discussed its broader applicability. This approach can be generalised  to other disciplines with different stability structures, as transition problems arise across a wide range of fields. \\

Looking forward, several promising directions warrant further investigation. Firstly, the current geometric framework can be generalised to accommodate alternative and more complex stability structures, such as transitions between multiple saddles (saddle-to-saddle cases) or networks embedded with heteroclinic channels. Secondly, while the current Large Deviation framework successfully determines the leading-order exponential scaling behaviours, a highly desirable and important extension would be to resolve the subexponential prefactor. Quantifying this prefactor as developed in alternative contexts \cite{Freddy2016,TB2023, Schorlepp2023, Heller2024} represents a key avenue to transition from scaling laws to absolute transition probabilities. Finally, it would be highly interesting to extend the Method of Division to systems driven by non-Gaussian noises, such as $t$-processes or L{\'e}vy stable processes. Such extensions, which feature heavy tails and are characterised by more than two cumulants, would significantly broaden the applicability of our computational framework to a wider class of non-Gaussian physical and engineering systems. \\

\begin{acknowledgments}
\noindent We are grateful to the reviewers and Ryan Deeley for useful comments.\\
For the purpose of open access, the authors have applied a Creative Commons Attribution (CC BY) license to any Author Accepted Manuscript version arising from this submission.\\
Jiayao Shao is supported by the Warwick Mathematics Institute Centre for Doctoral Training, and gratefully acknowledges funding from the University of Warwick.\\
Tobias Grafke acknowledges the support received from the EPSRC projects EP/T011866/1 and EP/V013319/1.\\
\end{acknowledgments}

\section*{Data Availability}
The data and computational codes that support the findings of this study are available within the article's Supplemental Material \cite{supp_material}.

\appendix
\section{Saddle Analysis}
\label{saddleanalysis}
\noindent The transformed linearised dynamics around the saddle are given by
\begin{align*}
    \dot{\tilde{\phi}}^1 &= \lambda_{+}\tilde{\phi}^1 + \tilde{\sigma}^1\eta, \\
    \dot{\tilde{\phi^i}} &= (\tilde{A_{s}}\tilde{\phi})^i + \tilde{\sigma}^i\eta \text{\quad \quad \quad   for $2 \leq i \leq n$},
\end{align*}
where the first component is the unstable direction. We aim to find an optimal $\eta$ such that for $0\leq t\leq\infty$, the dynamics satisfies $\tilde{\phi}^1(0)=\tilde{\phi}_T^1$ and $\tilde{\phi}^1(\infty)=0$.
By Duhamel's principle
\begin{align*}
     e^{-\lambda_+ t}\tilde{\phi}^1(t) =\tilde{\phi}^1(0) +\int_{0}^{t}e^{-\lambda_+ s}\tilde{\sigma}^1\eta(s)~ds . 
\end{align*}
Taking the limit $t\rightarrow\infty$ on both sides gives $\tilde{\phi}^1(0)=-\int_{0}^{\infty}e^{-\lambda_+ s}\tilde{\sigma}^1\eta(s)~ds$ since $\lim_{t\rightarrow \infty} e^{-\lambda_+ t}\tilde{\phi}^1(t) = 0$. Now, similar to the analysis for the sink, using the Lagrangian Multiplier method: 
\begin{align*}
G[\tilde{\phi^1_T},\eta,\beta] &= \frac{1}{2}\int^{\infty}_0 |\eta(s)|^2~ds + \beta(\tilde{\phi}^1_T-\tilde{\phi}^1(0)) \\
&=\frac{1}{2}\int^{\infty}_0 |\eta(s)|^2ds \\&+ \beta(\tilde{\phi}^1_T+\int_{0}^{\infty}e^{-\lambda_+ s}\tilde{\sigma}^1\eta(s)ds), \\
\frac{\delta G}{\delta\eta}&=\eta^T+\beta e^{-\lambda_+ s}\tilde{\sigma}^1 =0, \\
\eta &= -\beta e^{-\lambda_+ s}(\tilde{\sigma}^1)^T .
\end{align*}
Then the action around the saddle is
\begin{align*}
    \frac{1}{2}\int^{\infty}_0 |\eta|^2~dt &= \frac{\beta^2}{2}\int^{\infty}_0 e^{-\lambda_{+}s}(\tilde{\sigma}^1)(\tilde{\sigma}^1)^T e^{-\lambda_{+}s}~ds \\
    &= \frac{\beta^2}{2}q ,
\end{align*}
where $q=\frac{(\tilde{\sigma}^1)(\tilde{\sigma}^1)^T}{2\lambda_{+}}$. 

\noindent As under the optimal $\eta$, we have $\tilde{\phi}^1(0)=\tilde{\phi}^1_T$ with
\begin{align*}
    \tilde{\phi}^1_T &= \int_=^{\infty} e^{-\lambda_+ s}\tilde{\sigma}^1(\tilde{\sigma}^1)^T e^{-\lambda_+ s}\beta~ds \\
    &=q\beta.
\end{align*}
Then $\beta=q^{-1}\tilde{\phi}^1_T$ and the action takes the form $\frac{1}{2}q^{-1}(\tilde{\phi}^1_T)^2$. Moreover, the dynamics of the unstable component at time $t$ can be explicitly written as
\begin{align*}
    \tilde{\phi}^1(t) &=-\int_t^{\infty}e^{\lambda_+(t-s)}\tilde{\sigma}^1\eta~ds \\
    &= \int_t^{\infty}e^{\lambda_+(t-s)}\tilde{\sigma}^1(\tilde{\sigma}^1)^T e^{-\lambda_+ s}\beta~ds \\
    &= e^{-\lambda_+ t}\int_0^{\infty}e^{-\lambda_+\tau}\tilde{\sigma}^1(\tilde{\sigma}^1)^T e^{-\lambda_+\tau}\beta~d\tau\\
    &= \tilde{\phi}^1_T e^{-\lambda_+ t},
\end{align*} 
where we have substituted $s=t+\tau$ to shift the limit of the integral. This optimised $\eta$ fully determines the trajectory for the stable components in the following manner, for $2\leq i \leq n$:
\begin{align*}
    \dot{\tilde{\phi}^i} &=(\tilde{A_s}\tilde{\phi})^i+\tilde{\sigma}^i\eta \\
    &= (\tilde{A_s}\tilde{\phi})^i-\tilde{\sigma}^i(\tilde{\sigma}^1)^T e^{-\lambda_+ s}\tilde{\phi}^1_T q^{-1} \\
    &= (\tilde{A_s}\tilde{\phi})^i-\tilde{\sigma}^i(\tilde{\sigma}^1)^T q^{-1} \tilde{\phi}^1(t).
\end{align*}
Hence the linearised dynamics under control is given by
\begin{align*}
  \hat{A}=\begin{pmatrix}
  \begin{array}{c|c} 
  -\lambda_+ & 0 \\ 
  \hline 
  -q^{-1}\tilde{\sigma}^i(\tilde{\sigma}^1)^T  & \tilde{A_s} 
\end{array} 
\end{pmatrix}
\end{align*} 
where the $i^{th}$ entry of the first column is given by $-\tilde{\sigma}^i(\tilde{\sigma}^1)^T q^{-1}$ for $2\leq i \leq n$. The trajectories of the stable components can be derived as follows:
\begin{align*}
    & \dot{\tilde{\phi^i}} -(\tilde{A_s}\tilde{\phi})^i = -\tilde{\sigma}^i(\tilde{\sigma}^1)^T q^{-1} \tilde{\phi}^1(t), \\
    & \frac{d}{dt}(e^{-\tilde{A_s}t}\tilde{\phi}^i(t)) = -e^{-\tilde{A_s}t}\tilde{\sigma}^i(\tilde{\sigma}^1)^T q^{-1} e^{-\lambda_+ t}\tilde{\phi}^1(0), \\
    & e^{-\tilde{A_s}t}\tilde{\phi}^i(t)-\tilde{\phi}^i(0) \\&= -\int_0^t e^{-(\tilde{A_s}+\lambda_+ I)s}~ds~\tilde{\sigma}^i(\tilde{\sigma}^1)^T q^{-1} \tilde{\phi}^1(0), \\
    & \tilde{\phi}^i(t) = e^{\tilde{A_s}t}\tilde{\phi}^i(0) \\&+ (\tilde{A_s}+\lambda_+ I)^{-1}(e^{-\lambda_+ t}I-e^{\tilde{A_s}t})\tilde{\sigma}^i(\tilde{\sigma}^1)^T q^{-1} \tilde{\phi}^1(0),
\end{align*}
assuming that $-\lambda_+$ is not an eigenvalue of $\tilde{A}_s$. In the special case when $-\lambda_+$ is an eigenvalue of $\tilde{A_s}$, the integral involves powers of $t$ as well as exponentials; the result depends on the multiplicity of $-\lambda_+$ so we do not write it out. The point $\tilde{\phi}_T$ lies on the ellipsoid and is to be optimised for the overall problem. \\

\section{Lyapunov Equations}
\label{lyapunovequations}
\noindent In the linearisation analysis of the two endpoints, we employ two distinct forms of the Lyapunov equation. To clarify their physical significance, we derive both versions below and provide intuition regarding their respective roles in dynamical systems. The two versions are defined as follows:
\begin{align}
    A^TP+PA+Q&=0, \label{firstversion} \\
    AP+PA^T+Q&=0. \label{secondversion}
\end{align}

\noindent Given a linear system $\dot{x}=Ax$ and a symmetric matrix $Q$, (\ref{firstversion}) determines a symmetric matrix $P$ such that the rate of change of $x^TPx$ is $-x^TQx$.  In particular if $Q$ is chosen positive definite then $x^TPx$ decreases.  If $P$ turns out to be positive definite then this provides quantitative bounds on the stability of the origin. \\

\noindent Consider a linear dynamical system $\dot x = Ax$ where $A$ is stable. By defining a Lyapunov function $V(x)=x^TPx$ for some matrix $P$, its time derivative along the system trajectory is:
\begin{align*}
    \dot V(x) &= \dot x^T Px+x^TP\dot x, \\
    &= x^T(A^TP+PA)x, \\
    &= -x^TQx~.
\end{align*}
By choosing a positive definite matrix $Q$, $V(x)$ acts as an energy-like function that strictly decreases along the trajectory over time. Geometrically, the solution $P$ defines the level sets of the system energy. The ellipsoid $x^TPx\leq r^2$ characterises a forward-invariant subset from which the trajectory cannot escape, with the flow always directed inward. In our saddle analysis, we apply this form to the optimally controlled dynamics $\dot\phi=\hat{A}\phi$. By setting $Q=I$ (the identity matrix), we obtain a matrix $P=M$ that defines a forward-invariant ellipsoid $\phi^TM\phi\leq r^2$.\\

\noindent The version (\ref{secondversion}) describes the steady-state covariance of a linear stable stochastic dynamical system. Consider a linear system driven by white noise: $\dot x=Ax+\sigma\eta$, where $\eta$ is a zero-mean Gaussian white noise process and $A$ is stable. The steady-state covariance matrix $P$ can be expressed as the integral:
\begin{align*}
    P&=\int_{-\infty}^0e^{-As}\sigma\sigma^Te^{-A^Ts}~ds, \\
    &=\int_{-\infty}^0f(s)~ds.
\end{align*}
Let $f(s)=e^{-As}\sigma\sigma^Te^{-A^Ts}$ be the integrand. Consider the time derivative of $f(s)$:
\begin{align*}
    \frac{d}{ds}f(s) &= -Ae^{-As}\sigma\sigma^Te^{-A^Ts} +e^{-As}\sigma\sigma^Te^{-A^Ts}(-A^T), \\
    &= -Af(s) - f(s)A^T.
\end{align*}
Then
\begin{align*}
    \int_{-\infty}^0\frac{d}{ds}f(s)~ds &= \int_{-\infty}^0[-Af(s)-f(s)A^T] ~ds, \\
    [e^{-As}\sigma\sigma^Te^{-A^Ts}]_{-\infty}^0 &= -A\int_{-\infty}^0 f(s)~ds - \int_{-\infty}^0 f(s)~ds~A^T, \\
    \sigma\sigma^T-0 &= -AP-PA^T,
\end{align*}
by assuming $A$ stable. Rearranging gives the Lyapunov equation $AP+PA^T+\sigma\sigma^T=0$ with $Q=\sigma\sigma^T$. In our sink analysis, this form is used to characterise the stationary distribution of the stochastic process. Given the noise intensity (energy input) $Q=\sigma\sigma^T$, the solution $P$ describes the invariant measure of the system state under persistent fluctuations. \\

\section{Linearisation Error Estimation}
\label{linearisationerror}
\noindent The linearisation error can be estimated by representing the nonlinear dynamics as a small correction to the linearised dynamics and treating it to first order:
\begin{align*}
    b(\phi)=A\phi+\delta b(\phi).
\end{align*}
Then,
\begin{align*}
    \phi &= \bar{\phi}+\delta \phi ~, \\
    \eta &= \bar{\eta}+\delta \eta ~.
\end{align*}
We use $\bar{}$ on top of a variable to represent the zero-th order term and $\delta$ to the left of a variable to represent the first order correction. The dynamics becomes
\begin{align*}
    \dot\phi &= \dot{\bar{\phi}}+\dot{\delta\phi} \\
    &= A(\bar{\phi}+\delta\phi)+\delta b(\bar{\phi}+\delta\phi) +\sigma(\bar{\eta}+\delta\eta) ~.
\end{align*}
Rearranging via different orders gives:
\begin{align*}
    \text{zero-th order}:~&\dot{\bar{\phi}} = A\bar{\phi}+\sigma\bar{\eta}, \\
    \text{first order}:~&\dot{\delta\phi} = A\delta\phi +\delta b(\bar{\phi}) + \sigma \delta\eta
\end{align*}
and higher order terms. \\

\noindent We treat the case of a sink, so we are interested in the time interval from $-\infty$ to $0$. The perturbed objective function $J$ is 
\begin{align*}
    & \frac{1}{2}\int_{-\infty}^0(\bar{\eta}+\delta\eta)^T(\bar{\eta}+\delta\eta)~dt \\ 
    &+ \int_{-\infty}^0\langle(\bar{\mu}+\delta\mu),(\dot{\bar{\phi}}+\dot{\delta\phi})-A(\bar{\phi}+\delta\phi)-\delta b(\bar{\phi}+\delta\phi)\\&-\sigma(\bar{\eta}+\delta\eta)\rangle dt 
    + \langle\nu,\bar{\phi}(0)+\delta\phi(0)-\phi_0\rangle .
\end{align*}
Consider only the first order since the zero-th order is just the original problem:
\begin{align*}
    \delta J = &\int_{-\infty}^0 \bar{\eta}^T\delta\eta~dt + +\int_{-\infty}^0\delta\mu^T\dot{\bar{\phi}}-\delta\mu^TA\bar{\phi}-\delta\mu^T\sigma\bar{\eta}~dt \\
    &\phantom{==} \int_{-\infty}^0 \bar{\mu}^T\dot{\delta\phi}-\bar{\mu}^TA\delta\phi-\bar{\mu}^T\delta b(\bar{\phi})-\bar{\mu}^T\sigma\delta\eta~dt  \\
    =&\int_{-\infty}^0\bar{\eta}^T\delta\eta-\int_{-\infty}^0\dot\bar{\mu}^T\delta\phi~dt \\&- \int_{-\infty}^0(\delta\mu^TA\bar{\phi}+\delta\mu^T\sigma\bar{\eta})~dt
    +\int_{-\infty}^0 \dot{\delta\mu}^T\bar{\phi}~dt \\& -\int_{-\infty}^0 (\bar{\mu}^TA\delta\phi+\bar{\mu}^T\delta b(\bar{\phi})+\bar{\mu}^T\sigma\delta\eta)~dt.
\end{align*}
Taking variations: 
\begin{align*}
    \frac{\delta J}{\delta\bar{\phi}} &= - \bar{\mu}^T\delta\nabla b -\dot{\delta\mu}^T - \delta\mu^TA, \\
    \frac{\delta J}{\delta \bar{\eta}} &= \delta\eta-\delta\mu^T\sigma
\end{align*}
That is, at the optimum,
\begin{align*}
    \dot{\delta\mu} &= -A^T\delta\mu-\nabla \delta b^T\bar{\mu}, \\
    \delta\eta &= \sigma^T\delta\mu  
\end{align*}
The first order action is then
\begin{align*}
    \int_{-\infty}^0 \bar{\eta}^T\delta\eta~dt = \int_{-\infty}^0 \bar{\mu}^T\sigma\sigma^T\delta \mu~dt = \int_{-\infty}^0 \bar{\mu}^TC\delta\mu~dt.
\end{align*}
Also, the optimal first order trajectory satisfies
\begin{align*}
\dot{\delta\phi}=A\delta\phi+\delta b(\bar{\phi})+C\delta\mu
\end{align*}
where $C=\sigma\sigma^T$. For this forward equation, we know that $\delta\phi(-\infty)=0$ and we want $\delta\phi(0)=0$ so that the trajectory ends exactly at $\phi_0$ with no correction term. Then by Duhamel's principle,
\begin{equation}
    \delta\phi(0)=\int_{-\infty}^0e^{-As}(\delta b(\bar{\phi})+C\delta\mu)~ds = 0
\end{equation}
which gives
\begin{align*}
    \int_{-\infty}^0e^{-As}\delta b(\bar{\phi})~ds = -\int_{-\infty}^0 e^{-As}C\delta\mu~ds \tag{$\star$}
\end{align*}
Note that at zero-th order under optimised condition,
\begin{align*}
    \dot{\bar{\phi}} &= A\bar{\phi}+C\bar{\mu}, \\
    \dot{\bar{\mu}} &= -A^T\bar{\mu} .
\end{align*}
The forward $\phi$-equation terminates at $\bar{\phi}(0)$ which provides the boundary condition for the backward $\mu$-equation at $\bar{\mu}(0)$:
\begin{align*}
    \bar{\phi}(0)&=\int_{-\infty}^0e^{-As}C\bar{\mu}(s)~ds\\
    &=\int_{-\infty}^0 e^{-As}Ce^{-A^Ts}\bar{\mu}(0)~ds = Q\bar{\mu}(0),
\end{align*}
where $Q=\int_{-\infty}^0e^{-As}Ce^{-A^Ts}$ (this just reproduces the same $Q$ as in the sink analysis section). Thus the boundary condition for $\bar{\mu}$ at $t=0$ is
\begin{align*}
    \bar{\mu}(0) = Q^{-1}\phi_0
\end{align*}
The first order action can then be reformulated as
\begin{align*}
    &\int_{-\infty}^0 \bar{\mu}(s)^TC\delta\mu(s)~ds \\&= \int_{-\infty}^0 \bar{\mu}(0)^Te^{-As}C\delta\mu(s)~ds \\
    &=  \phi_0^TQ^{-1}\int_{-\infty}^0e^{-As}C\delta\mu(s) ~ds \\
    &= -\phi_0^TQ^{-1}\int_{-\infty}^0e^{-As}\delta b(Qe^{-A^Ts}Q^{-1}\phi_0) ~ds
\end{align*}
using ($\star$) and then $\bar{\phi}(s)=Qe^{-A^Ts}Q^{-1}\phi_0$. This action correction is of $O(\phi_0^3)$. One can compare it with the zero-th order action under linearisation $\frac{1}{2}\bar{\phi}_0^TQ^{-1}\phi_0$ and obtain an order of magnitude estimate for the integral:
\begin{align*}
    \frac{\delta S}{\bar{S}} \approx \frac{2|\delta b(\phi_0)|}{c|\phi_0|} ,
\end{align*}
where $c$ is the slowest contraction rate for $A$. \\

\noindent The above derivation is applicable to generic cases like the sink analysis in our method. It is similar for the saddle analysis but coordinate transformation should be taken into consideration (assume the saddle is at the origin):
\begin{align*}
    U^{-1}\phi &= U^{-1}\bar{\phi} + U^{-1}\delta\phi,\\
    \tilde{\phi} &= \tilde{\bar{\phi}} + \tilde{\delta \phi} ,
\end{align*}
where $U$ is the coordinate transformation matrix defined in equation (\ref{transformationmatrix}) and the dynamics is 
\begin{align*}
    \tilde{b}(\tilde{\phi}) &= U^{-1} b(U\tilde{\phi}) = \tilde{A} \tilde{\phi} + \tilde{\delta b}(\tilde{\phi}),\\
    \tilde{\delta b} &= U^{-1}\delta b (U\tilde{\phi}) ,
\end{align*}
where $\tilde{A}$ is in block form with the unstable component separated from the stable components. The first order action in transformed coordinates is then:
\begin{align*}
    -\frac{\tilde{\phi}^1_T}{q}\int_0^{\infty}e^{-\lambda_+ s}\tilde{\delta b}^1(\tilde{\phi}(s))~ds
\end{align*}
where the dynamics of $\tilde{\phi}(s)$ is derived in Appendix \ref{saddleanalysis}.

\section{Discretised Gradient}
\label{discretisation}
\noindent The objective cost function $J$ in its discrete version (using forward Euler method to discretise the integrals) takes the form
\begin{align*}
    J[\phi,\eta,\mu,\nu,\beta,\lambda]= &\frac{1}{2}\sum_{i,j=1}^{n}\phi_0^i Q^{-1}_{ij}\phi_0^j + \frac{1}{2}\sum_{k=1}^{N-1} \sum_{i}^n (\eta_k^i)^2\Delta t \\&+ \frac{1}{2}q^{-1}(\tilde{\phi}(T)^1)^2 + \sum_{i=1}^{n}\nu_i (\phi_0^i -\phi(0)^i) \\&+ \sum_{k=1}^{N-1}\sum_{i,j=1}^n\mu_k^i(\dot\phi_k^i-b(\phi_k^i)-\sigma_{ij}\eta_k^j) \Delta t \\
    &+ \beta(\sum_{i,j=1}^n\tilde{\phi}(T)^i M_{ij} \tilde{\phi}(T)^j - r^2) \\&+ \lambda(\sum_{i,j=1}^n\tilde{\phi}(T)^i M_{ij} \tilde{\phi}(T)^j - r^2)^2
\end{align*}
where $n$ is the dimension of the SDE so $\phi \in \mathbb{R}^n$ and the total transition time $T$ is discretised into $N$ points such that $\phi_1=\phi(t=0)$ and $\phi_N=\phi(t=T)$. The Lagrange multiplier $\mu$ is correspondingly discretised into $N-1$ points. The reason that $\mu$ has one point less than $\phi$ will be explained later. Also note that, under the transformation, $\tilde{\phi}_N = U^{-1}(\phi_N -\textbf{s})$ where $\textbf{s}$ is the saddle. Then the above equation is equivalent to
\begin{align*}
     J[\phi]= &\frac{1}{2}\sum_{i,j=1}^{n}\phi_0^i Q^{-1}_{ij}\phi_0^j + \frac{1}{2}\sum_{k=1}^{N-1} \sum_{i}^n (\eta_k^i)^2\Delta t \\&+ \frac{1}{2}q^{-1}((\tilde{\phi}_N)^1)^2 + \sum_{i=1}^{n}\nu_i (\phi_0^i -\phi_1^i) \\&+ \sum_{k=1}^{N-1}\sum_{i,j=1}^n\mu_k^i(\frac{\phi^i_{k+1}-\phi^i_{k}}{\Delta t}-b(\phi_k^i)-\sigma_{ij}\eta_k^j) \Delta t \\
    &+ \beta(\sum_{i,j=1}^n\tilde{\phi}_N^i M_{ij} \tilde{\phi}_N^j - r^2) \\&+ \lambda(\sum_{i,j=1}^n\tilde{\phi}_N^i M_{ij} \tilde{\phi}_N^j - r^2)^2
\end{align*}
Varying $J$ with respect to $\phi_0$ and $\eta$ gives:
\begin{align*}
    \frac{\partial J}{\partial\phi_0^i} &= \sum_{j=1}^n Q_{ij}\phi_0^j +\nu_i, \\
    \frac{\partial J}{\partial\eta_k^i} &= \eta_k^i - \sum_{j=1}^n \sigma_{ij}\mu_k^j
\end{align*}
where the $\dot\phi$ is discretised via forward Euler method. Varying $J$ with respect to $\phi$ gives:
\begin{align*}
    \frac{\partial J}{\partial\phi_1^i} &= -\nu_i - \mu_1^i - \mu_1^i\nabla b(\phi_1^i)^T\Delta t, \\
    \frac{\partial J}{\partial\phi_k^i} &= \mu_{k-1}^i -\mu_{k}^i - \mu_k^i\nabla b(\phi_k^i)^T\Delta t 
\end{align*}
For $\phi$ and $\mu$, the superscript indicates the dimension from $1$ to $n$ and the subscript indicates the time index from $0$ to $N$. Here note that $$\tilde{\phi}_N^T M \tilde{\phi}_N = (\phi_N-\textbf{s})^T U^{-T}MU^{-1}(\phi_N-\textbf{s}),$$ where $\textbf{s}$ is the saddle, so varying $J$ with respect to $\phi_N$ gives:
\begin{align*}
    \frac{\partial J}{\partial\phi_N} &= \mu_{N-1}+\frac{(l^T(\phi_N-\textbf{s}))l}{q} + 2(U^{-T}MU^{-1})(\phi_N-\textbf{s})\beta \\
    &+ 4\lambda(U^{-T}MU^{-1})(\phi_N-\textbf{s})(\tilde{\phi}_N ^T M \tilde{\phi}_N -r^2)).
\end{align*}
Thus, in optimal conditions,
\begin{align*}
    &\nu = -\mu_1 - \mu_1\nabla b(\phi_1)^T\Delta t, \\
    &\mu_{i-1} = \mu_{i} + \mu_i\nabla b(\phi_i)^T\Delta t, \\
    &\mu_{N-1} = -\frac{(l^T(\phi_N-\textbf{s}))l}{q} \\
    &- 2(U^{-T}MU^{-1})(\phi_N-\textbf{s})(\beta + 2\lambda(\tilde{\phi}_N ^T M \tilde{\phi}_N -r^2)).
\end{align*}
These $\mu$ equations describe the backward Hamilton's equation where $\phi_N$ provides the boundary information for $\mu_{N-1}$ and, iteratively, $\mu_k$ is determined by $\mu_{k+1}$ and $\phi_{k+1}$. Therefore, we would only obtain $N-1$ points for $\mu$ and that $\mu_1$ and $\phi_1$ together determines $\nu$ so that the gradient $\frac{\partial J}{\partial\phi_0}$ is interpretable. \\

\noindent Varying $J$ with respect to $\mu$ gives:
\begin{align*}
    \frac{\partial J}{\partial\mu^i_k} = \dot\phi_k^i-b(\phi_k^i)-\sum_{j=1}^n\sigma_{ij}\eta_k^j
\end{align*}
that is the forward Hamilton's equation when $\frac{\partial J}{\partial\mu^i_k}=0$. We can interpret $\dot\phi$ by finite difference such that
\begin{align*}
    \phi_{k+1}^i = \sum_{j=1}^n(b(\phi_k^i)+\sigma_{ij}\eta_k^j)\Delta t + \phi_k^i
\end{align*}

\section{Inner Product for the Gradient}
\label{gradientnorm}
\noindent In general, an inner product on a vector space $V$ is defined via a positive-definite matrix $g$: $\langle u,v\rangle=u^i g_{ij} v^j$ for $u$, $v \in V$. Consider a smooth function $f: V \rightarrow \mathbb{R}$. The gradient of $f$ at a general point $x_0$ is the vector $\nabla f$ defined by 
\begin{align*}
    \langle v,\nabla f\rangle &= df(v) \text{\quad \quad } \forall v \in V \\
    v^i g_{ij} \nabla^j f &= v^i \frac{\partial f}{\partial x^i} \text{\quad \quad } \forall v \in V
\end{align*}
as $df=\frac{\partial f}{\partial x^i}dx^i$. Thus
\begin{align*}
    \nabla^j f = g^{ji}\frac{\partial f}{\partial x^i}
\end{align*}
where $g^{ji}$ is the inverse matrix of $g_{ij}$. \\

\noindent With our model, define an inner product on the vector space $V=(\phi_0, \eta)$ by
\begin{align*}
    \langle (\phi_0,\eta),(\tilde{\phi_0},\tilde{\eta})\rangle = \phi_0\tilde{\phi_0} + K\sum_{i=1}^{N-1}\eta_i\tilde{\eta_i}\Delta t
\end{align*}
for some constant $K$ (for simplicity let $K=1$). Then the inner product is defined by a diagonal matrix $G$ with $g_{11}=1$ and $g_{ii}=\Delta t$ for $2 \leq i$. We have the variation $dJ$:
\begin{align*}
    dJ = &(Q\phi_0 + \nu)d\phi_0  + \sum_{i=1}^{N-1}(\eta_i-\sigma^T\mu_i)\Delta t d\eta_i   
\end{align*}
Then, the gradient of $J$ is given by
\begin{align*}
    \nabla J &= G^{-1}dJ \\
    &= (Q\phi_0 + \nu, \eta_1-\sigma^T\mu_1,...,\eta_{N-1}-\sigma^T\mu_{N-1}).
\end{align*}
Therefore, the gradient of $J$ on $\phi_0$ is
\begin{align*}
    Q\phi_0 + \nu
\end{align*}
and the gradient of $J$ on $\eta$ is 
\begin{align*}
    \eta-\sigma^T\mu.
\end{align*}

\noindent In practice, we update $\eta$ and $\phi_0$ via nonlinear conjugate gradient descent \cite{NumericalOptimisation} with Fletcher-Reeves coefficients \cite{NCGbeta}: 
\begin{align*}
    \gamma &= \frac{(\nabla J)_k^T (\nabla J)_k}{(\nabla J)_{k-1}^T (\nabla J)_{k-1}}, \\
    CG_{k+1} &= -(\nabla J)_k + \gamma\cdot CG_k, \\
    (\nabla J)_{k+1} &= (\nabla J)_k + \alpha\cdot CG_{k+1},
    \end{align*}
where $CG_i$ denote the conjugate gradient for the $i^{th}$ iteration and $\alpha$ is the step size determined by line search.\\

\section{Linearised Ship Dynamics}
\label{shipdynamics}
\noindent The linearised ship dynamics is analysed in this section. Consider only the ship components with damping, but no forcing. The linearised dynamics is given by
\begin{align*}
\label{equ_linearisation}
    \nabla b =
\begin{pmatrix}
    0 & 1 & 0 & 0\\
    -\frac{h}{m} & -k_1 & \frac{h\gamma}{m}\theta & 0\\
    0 & 0 & 0 & 1\\
    \frac{h\gamma}{I}\theta & 0 & (\frac{c}{I\theta_c^2}-\frac{h\gamma^2}{2I})3\theta^2 + (\frac{h\gamma}{I}z-\frac{c}{I}) & -k_2
\end{pmatrix} .
\end{align*} 
Then the linearised dynamics around the sink $\textbf{0}\in\mathbb{R}^4$ is
\begin{align*}
A_{-\infty} = 
\begin{pmatrix}
    0 & 1 & 0 & 0 \\
    -\frac{h}{m} & -k_1 & 0 & 0\\
    0 & 0 & 0 & 1 \\
    0 & 0 & -\frac{c}{I} & -k_2 
\end{pmatrix}.
\end{align*}
The eigenvalues are 
\begin{align*}
\lambda_1 &= \frac{-k_1 + \sqrt{k_1^2 - \dfrac{4h}{m}}}{2}, \quad \lambda_2 = \frac{-k_1 - \sqrt{k_1^2 - \dfrac{4h}{m}}}{2}, 
\\
\lambda_3 &= \frac{-k_2 + \sqrt{k_2^2 - \dfrac{4c}{I}}}{2}, \quad \lambda_4 = \frac{-k_2 - \sqrt{k_2^2 - \dfrac{4c}{I}}}{2}.
\end{align*}
If there is no damping, i.e. $k_1=k_2=0$, then we would have completely imaginary eigenvalues $\pm\sqrt{\frac{h}{m}}$ and $\pm\sqrt{\frac{c}{I}}$. These correspond to the natural oscillation frequencies of the heave and the roll. We have damped oscillatory behaviour around the sink if $0<k_1<\sqrt{\frac{4h}{m}}$ and $0<k_2<\sqrt{\frac{4c}{I}}$. \\

\noindent The linearised dynamics around the saddle $(\frac{1}{2}\gamma\theta_c^2,0,\pm\theta_c,0)$ are given by
\begin{align*}
A_{\infty} = 
\begin{pmatrix}
    0 & 1 & 0 & 0\\
    -\frac{h}{m} & -k_1 & \frac{h\gamma \theta_c}{m} & 0\\
    0 & 0 & 0 & 1\\
    \frac{h\gamma \theta_c}{I} & 0 & (\frac{c}{I\theta_c^2}-\frac{h\gamma^2}{2I})3\theta_c^2-\frac{c}{I} & -k_2\\
\end{pmatrix}.
\end{align*}
This equilibrium point has a pair of complex eigenvalues with negative real part and a pair of real eigenvalues with one positive and one negative. The saddle has a one-dimensional unstable manifold and three-dimensional stable manifold. \\

\section{Noise Intensity}
\label{noiseintensity}
\noindent We wish to quantify the intensity of random forcing functions in a way that allows sensible comparison of the effects on a system as the probability distribution of the forcing is changed.  For example, how does the capsize rate vary as the characteristics of a filtered white noise are varied?  The variance of the white noise is not necessarily the appropriate quantification of the filtered noise intensity. \\

\noindent First, choose an inner product on the state space so that the size of a force $\zeta$ at a given time is quantified by $\sqrt{ \zeta^T K^T K \zeta}$ with $K$ invertible (this allows for example to compare the effects of different components of the force, some of which might be linear forces, others torques).  In our ship problem we have already non-dimensionalised the ship variables so we just take $K$ to be the identity matrix. \\

\noindent For a white noise forcing $\sqrt{\epsilon} \sigma \eta$, with $\eta$ a standard multidimensional white noise, 
a suitable quantification of its intensity is then the $\ell_2$-norm of $ \epsilon K\sigma \sigma^T K^T$. The matrix $\epsilon \sigma \sigma^T$ can be interpreted as the integrated autocorrelation function of the forcing, which is delta-correlated in this white case.  This suggests the following quantification of the intensity of a general random forcing function. \\

\noindent{\bf Definition}: The {\em intensity} $\mathcal{I}_\zeta$ of a random forcing function $\zeta$ on a system is the $\ell_2$-norm of $KJK^T$, where $J$ is the integrated autocorrelation function of $\zeta$. \\

\noindent Recalling that the $\ell_2$-norm of a symmetric matrix is its largest eigenvalue in absolute value, and that an integrated autocorrelation matrix $J$ is always positive semi-definite, the intensity can alternatively be defined as the largest eigenvalue of $KJK^T$. \\

\noindent If $\zeta = B \xi$ with $\xi$ a filtered white noise then the autocorrelation function of $\xi$ is given in numerous texts (e.g. \citet{caines1988}, \citet{StochasticMethodsBook}). In particular, if $\xi$ is 
the solution $\xi(t) = \int_{-\infty}^t e^{A(t-s)} \sqrt{\epsilon} \sigma \eta(s)\, ds$ of 
\begin{align*}
    \dot\xi = A\xi+\sqrt{\epsilon}\sigma\eta~,~
\end{align*}
for an asymptotically stable matrix $A$, then
the autocorrelation function $C_\xi$ of $\xi$ is $C_\xi(\tau) = \Sigma e^{A^T\tau}$ for $\tau>0$, $e^{-A\tau} \Sigma$ for $\tau<0$, where $\Sigma$ is the solution of the Lyapunov equation $A\Sigma + \Sigma A^T + \epsilon \sigma\sigma^T=0$.
It follows that its integral is
$$\int_{-\infty}^\infty C_\xi(\tau)\, d\tau = -\Sigma A^{-T} - A^{-1}\Sigma.$$
From the Lyapunov equation, this can be written as $$\epsilon A^{-1}\sigma \sigma^T A^{-T}.$$
Then 
$$J = \epsilon B A^{-1} \sigma \sigma^T A^{-T} B^T.$$
Define
$$\lambda_{\max} = \mbox{largest eigenvalue of }  KBA^{-1}\sigma \sigma^T A^{-T} B^T K^T .$$
Then the intensity $\mathcal{I}_\zeta$ is $\epsilon \lambda_{\max}$.
$\lambda_{\max}$ can equivalently be defined as the square of the largest singular value of the matrix $KBA^{-1}\sigma$. \\

\noindent Large deviation theory gives the capsize probability rate as $r \asymp \exp{-S/\epsilon}$ for the minimum $S$ of the action.  We rewrite this as $\exp(-S \lambda_{\max}/\mathcal{I})$. So to make fair comparisons of the minimum action as filter parameters are varied, for equal forcing intensity, we plot the scaled action $S \lambda_{\max}$.

\bibliography{reference}

\end{document}